\documentclass[]{jfm}

\usepackage[justification=justified,singlelinecheck=false]{caption}

\usepackage{amsmath}
\usepackage{amssymb}
\usepackage{graphicx}
\usepackage{newtxtext}
\usepackage{newtxmath}
\usepackage{natbib}
\usepackage{hyperref}
\hypersetup{
    colorlinks = true,
    urlcolor   = blue,
    citecolor  = black,
}

\newcommand{\RomanNumeralCaps}[1]

\newcommand{\bp}{\boldsymbol{p}}
\newcommand{\bx}{\boldsymbol{x}}

\DeclareMathOperator*{\argmax}{arg\,max}

\title{Dynamics and clustering of sedimenting disc lattices}

\author{Harshit Joshi\aff{1}
\corresp{\email{harshit.joshi@icts.res.in}},
  Rahul Chajwa\aff{2},
  Sriram Ramaswamy\aff{3,1},
  Narayanan Menon\aff{4}
 \and Rama Govindarajan\aff{1}}

\affiliation{\aff{1}International Center for Theoretical Sciences, Bengaluru, 560 089
\aff{2}Department of Bioengineering, Stanford University, Stanford, California 94305, USA
\aff{3} Centre for Condensed Matter Theory, Department of Physics, Indian Institute of Science, Bengaluru, Karnataka 560 012
\aff{4} Department of Physics, University of Massachusetts, Amherst, Massachusetts 01003, USA
}

\begin{document}
\maketitle

\begin{abstract}
Uniform arrays of particles tend to cluster as they sediment in viscous fluids. Shape anisotropy of the particles enriches these dynamics by modifying the mode-structure and the resulting instabilities of the array. A one-dimensional lattice of sedimenting spheroids in the Stokesian regime displays either an exponential or an algebraic rate of clustering depending on the initial lattice spacing \citep{chajwa2020waves}. This is caused by an interplay between the Crowley mechanism which promotes clumping, and a shape-induced drift mechanism which subdues it.  We theoretically and experimentally investigate the sedimentation dynamics of one-dimensional lattices of oblate spheroids or discs and show a stark difference in clustering behaviour: the  Crowley mechanism results in clumps comprised of several spheroids, whereas the drift mechanism results in
pairs of spheroids whose asymptotic behavior is determined by pair-hydrodynamic interactions. 
We find that a Stokeslet, or point-particle, approximation is insufficient to accurately describe the instability and that the corrections provided by the first-reflection are necessary for obtaining some crucial dynamical features. As opposed to a sharp boundary between exponential growth and neutral eigenvalues under the Stokeslet approximation, the first-reflection correction leads to exponential growth for all initial perturbations, but far more rapid algebraic growth than exponential growth at large lattice spacing $\Tilde{d}$. For discs with aspect ratio $0.125$, corresponding to the experimental value, the instability growth rate is found to decrease with increasing lattice spacing $\Tilde{d}$, approximately as $\Tilde{d}^{ -4.5}$, which is faster than the $\Tilde{d}^{-2}$ for spheres \citet{crowley1971viscosity}. It is shown that the first-reflection correction has a stabilizing effect for small lattice spacing and a destabilizing effect for large lattice spacing.
Sedimenting pairs predominantly come together to form an inverted `T', or $\perp$, which our theory accounts for through an analysis that builds on  \citet{koch1989instability}. This structure remains stable for a significant amount of time.
\end{abstract}



\section{Introduction} \label{intro}
Particle sedimentation through fluid is ubiquitous in natural and industrial processes, such as the settling of clay particles on to a riverbed, precipitates in a chemical reaction, diatoms in
pelagic algal blooms and ice crystals in cirrus clouds. 
Collections of sedimenting spherical particles have a tendency to clump \citep{crowley1971viscosity}, leading to the formation of large aggregates and thence to increased sedimentation rates. In an otherwise quiescent fluid clumping is a consequence of hydrodynamic interactions between settling particles, which in turn depends on 
the shape and relative orientation of particles and inter-particle separations. 

Most particles in nature are not perfect spheres, and shape anisotropy in sedimenting particles 
leads to qualitatively new dynamical features. 
Spheroids offer a simple model to describe departures in particle shape from sphericity, and exhibit much richer sedimentation dynamics than spheres, both because isolated spheroids can drift horizontally and pairs are coupled through their orientational and translational degrees of freedom \citep{Jung2006, chajwa2019kepler, jeffery1922motion, kim1985sedimentation}. These new elements in the dynamics can affect particle collision rates and their tendency to form aggregates.

Spheroidal particles range from disc-like (oblate) to rod-like (prolate). The Stokesian sedimentation of 
a suspension of slender fibres 
 has been studied experimentally \citep{Guazzelli2011} and numerically \citep{MACKAPLOW1998, Gustavsson2009} 
with a focus on the dependence of velocity fluctuations and structure factor on 
length scale and particle concentration. 
In very low volume fractions, these studies report large-scale inhomogeneities, with dense patches, termed streamers \citep{Metzger2005}, 
in which the mean particle orientation aligns with gravity, harbouring fast-settling clusters of fibres \citep{Gustavsson2009} that spontaneously form and break \citep{Guazzelli2011}.
The formation of clusters and streamers from an initially homogeneous suspension
occurs in a highly nonlinear regime; how this nonlinear state is arrived at from linear
instability \citep{koch1989instability} and what sets the typical size of clusters, is
unresolved \citep{Guazzelli2011}. 
Studies of sedimenting lattices of particles \citep{crowley1971viscosity, LR1997, chajwa2020waves} shed light on how particle-level interactions yield long-wavelength collective modes, and offer a simple setting in which to study the clustering that results from perturbations about 
a well-defined reference configuration.

In this article, we investigate numerically and experimentally the sedimentation of one-dimensional arrays of oblate spheroids in the limit of negligibly small Reynolds number and Stokes number. Going beyond the linear stability analysis of \citet{chajwa2020waves}, we elucidate the detailed dynamics at the level of individual particles, with emphasis on clustering due to hydrodynamic interactions. We show that, unlike spheres, discs display two qualitatively different behaviours of clustering depending on their initial spacing. At small initial spacing clustering is dictated by the Crowley mechanism, where clumps typically form at valleys in the initial perturbation and consist of several particles.
At larger spacing the drift mechanism becomes dominant, and in the non-linear regime, instead of clumps we obtain attracting pairs typically forming away from valleys in the initial perturbation pattern, which then fall together in a `$\perp$' configuration (see figure \ref{fig:lowQClump}). Two distinct forms of perturbation growth, arising from competing mechanisms, were analyzed earlier \citep{chajwa2020waves} in the linearized dynamics.
\citet{chajwa2020waves} assume a point particle approximation for hydrodynamic interactions, and obtain a sharp boundary between unstable and neutrally stable configurations. Here, we improve the approximation by correcting the boundary conditions at each particles by the method of reflections, and find that incorporating corrections of sub-leading order in $a/r$ (where $a$ is a typical particle size and $r$ the typical separation between particles), makes the system linearly unstable for all initial spacings and wavenumbers. 
For larger initial inter-particle spacing the exponential growth rates are extremely small, while transient algebraic growth of perturbations, driven by drift, are not, and dictate clustering into pairs at the antinodes, rather than clumps at the nodes. 

\begin{figure}
\centering
\includegraphics[width=0.8\textwidth]{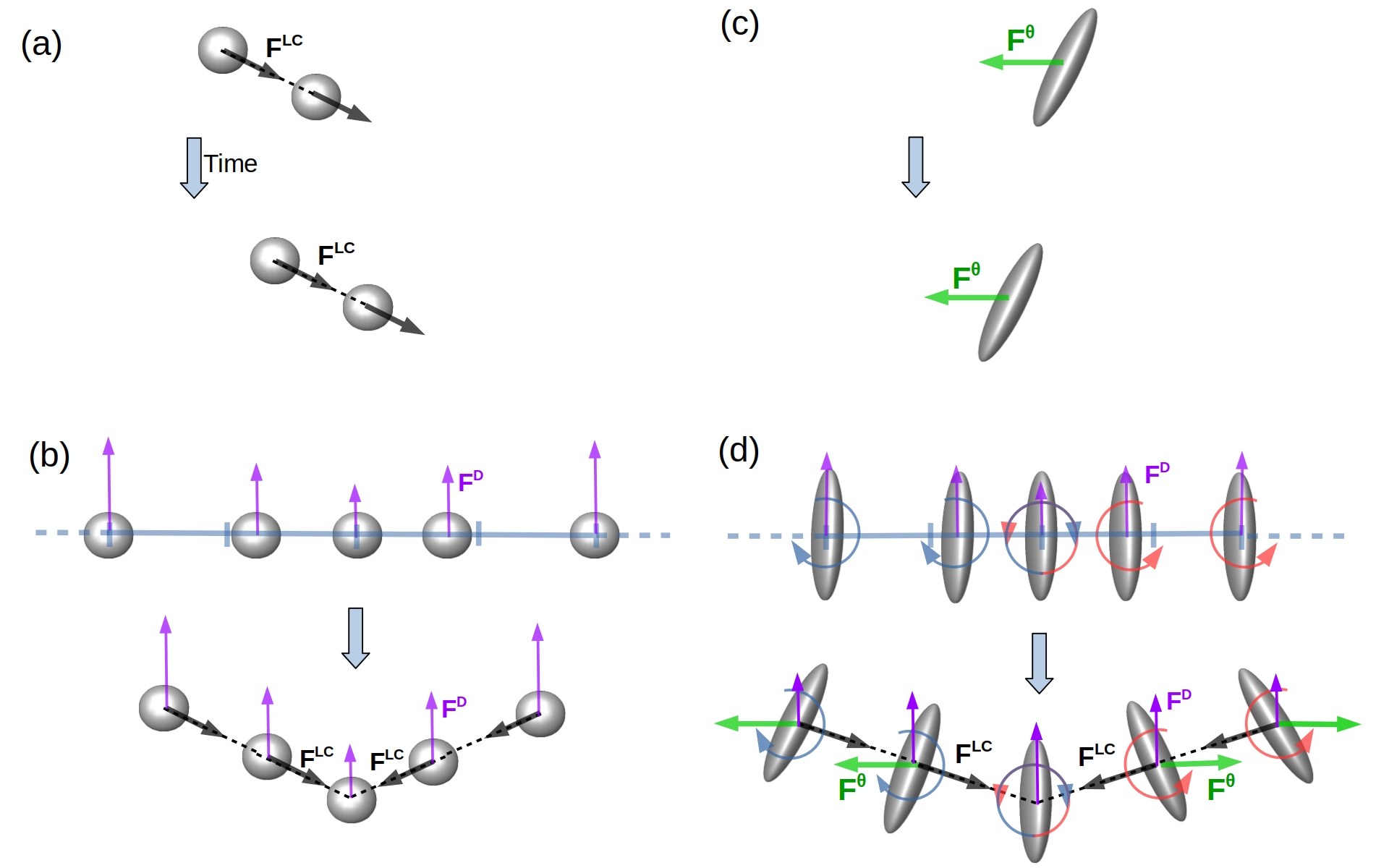}
\caption{\label{fig:mechanisms} Schematic showing how the drift mechanism competes with the Crowley mechanism. Adapted from \citet{chajwa2020waves}. The Crowley mechanism due to hydrodynamic interactions is operative irrespective of the shape of the particles and tends to form clumps at  valleys or at density nodes. In addition, spheroids, as they sediment, experience a horizontal drift as a consequence of their orientation, as shown by the green arrows. This orientation-dependent drift mechanism, operating at the level of individual particles, can suppress the Crowley instability. }
\end{figure}
 
This paper is organised as follows: the experimental setup and the theoretical approach we adopt are described in sections 2 and 3 respectively. We write down the evolution equations for the positions and orientations of the discs, emerging from the velocities and gradients of the Stokes flows generated by their gravitational force densities.
The method of reflections is used to obtain the background disturbed flow produced by each spheroid. 
Near-contact dynamics without considering lubrication forces can lead to overlapping spheroids. We adopt a numerical procedure to avoid overlaps 
with a simple model for the lubrication regime.
In section 4 we validate the governing equations by comparing them to our experimental results for the dynamics of a pair of discs. An interesting rocking motion was noticed during our validation exercises: two discs stacked in an `=' formation represent an equilibrium system which is neutrally stable to periodic perturbations. The stability analysis is conducted to understand the rocking dynamics. This exercise also brings out the limitations of a point-force approximation, and justifies the use of the first-reflection approximation in our theory. Section 5 describes the clustering behaviour of perturbations of a one-dimensional array of sedimenting oblate spheroids. The clustering of spheroids is analyzed in two regimes dominated by two different mechanisms: the Crowley mechanism and a drift mechanism. The two regimes are explored  numerically and we construct statistical measures that allow us to distinguish the two regimes. Section 6 is devoted to the understanding of the long-lived inverted `T' or `$\perp$' structures that form in the algebraic growth regime. We summarise our work and discuss the future avenues that emerge out of it in section 7.

\section{Experiments}
\begin{figure}
\centering
\includegraphics[width=1.0\textwidth]{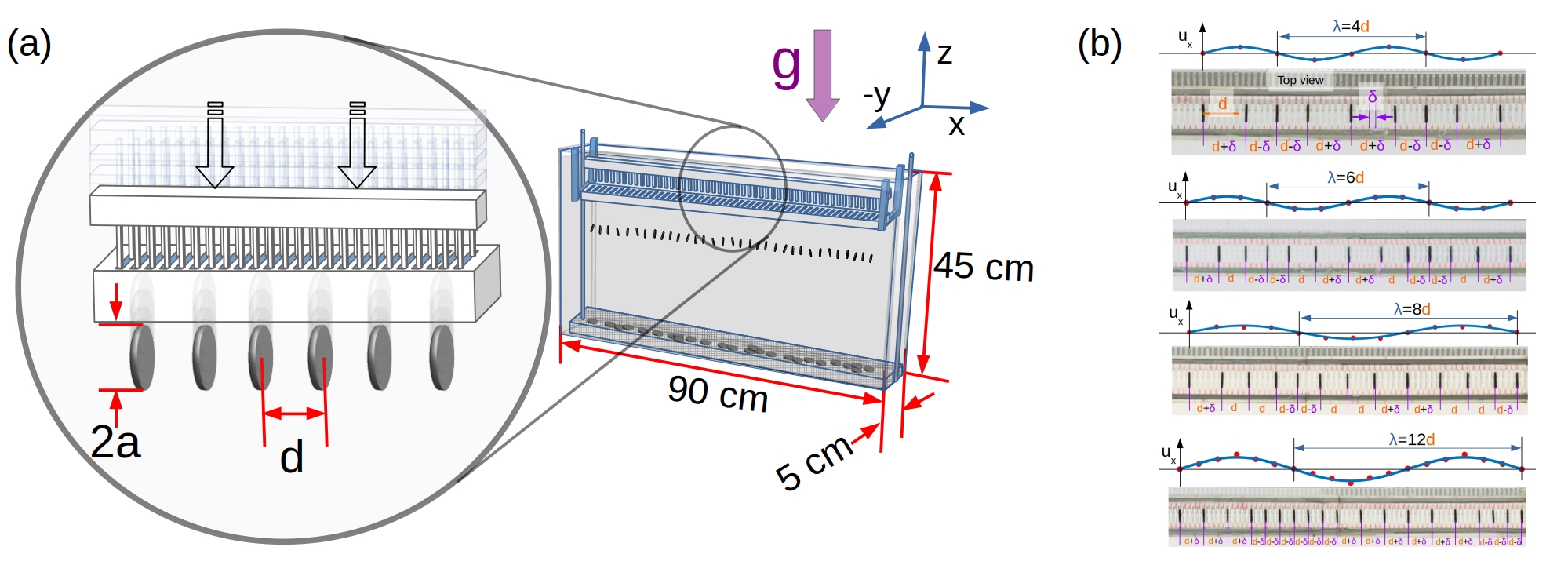}
\caption{\label{fig:setup} \textbf{Experimental setup:} (a) Shows the quasi-two-dimensional geometry of the container with gravity axis pointing along $-\hat{z}$. The mechanism simultaneously releases discs with controlled lattice spacing and perturbation wave-number. (b) the bottom-view of the release mechanism showing the array of discs moments before they were released in the fluid. The initial spacing are chosen to control the dimensional lattice spacing $d$ and the dimensional perturbation wave-number $2\pi/\lambda$.
}
\end{figure}

The sedimentation experiments were performed in a quasi-2D slab geometry (see Fig. \ref{fig:setup}) with height $45$ cm, width $90$ cm and thickness $5$ cm filled with silicone oil with density $0.98$ g/cm$^{3}$ and kinematic viscosity $5000$ cSt. Discs of diameter $2a = 8$ mm
and thickness $2b=1$ mm were 3D-printed with resin of density $1.164$ g/cm$^{3}$. These discs are modelled in our theory as oblate spheroids of aspect ratio $b/a=0.125$. 
The Reynolds number using the length scale set by the particle size was measured to be $\sim 10^{-4}$, ensuring that we are in a Stokesian regime. The quasi-2D geometry of the container has two effects: it leads to a modest reduction of the single-particle sedimentation speed, and it cuts off hydrodynamic interactions with neighbours beyond a length-scale set by the thickness.

Releasing multiple particles in a viscous fluid presents a challenge in synchronizing release times and controlling initial orientations, as pointed out by Jung et al. 2006, Chajwa 2020, and described in \citet{rahulThesis}.
Our release mechanism consists of an array of slots where discs can be selectively placed to define the initial (dimensional) lattice spacing $d$ and dimensional perturbation wave-number $2 \pi/\lambda$ [see figure \ref{fig:setup}]. The release mechanism is centered along the thickness of the slab and immersed in the fluid to remove bubbles attached to the discs. The discs are ejected from the slots simultaneously using a `comb' that inserts into the slots. This setup allows us to study the dependence of the particle dynamics on $d$ and $\lambda$.  A D-SLR camera is used to capture images of the discs every 3 seconds. The images are thresholded, and an ellipse is fit to each disc to determine its orientation and the location of its centroid. In the later stages, the thresholded images had to be manually segmented to separate closely-clustered discs.

\section{Governing equations}\label{sec:govEqns}

Our system consists of hydrodynamically interacting oblate spheroids (discs) labeled by $\alpha \in \{1, ...\, N_d\}$ sedimenting under gravity in an otherwise quiescent unbounded fluid. The $\alpha$th disc is described by  semi-major and semi-minor axes
$a$ and $b$, respectively, a centre position $\boldsymbol x_\alpha$, orientation unit vector $\boldsymbol p_\alpha$ 
which points along the symmetry axis of the disc, and instantaneous velocity $\boldsymbol{V}_\alpha$. 
For particle Reynolds number $Re_p\ll 1$, 
every disc generates a disturbance field that satisfies the Stokes equation
\citep{kim1985sedimentation,kim2013microhydrodynamics}. Moreover, when the particle relaxation time scale $\tau_p$ is much smaller than the time 
it takes to fall through its own length, 
the Stokes number $St \equiv (\tau_p V_\alpha/a) \ll 1$ for each disc, so that its buoyancy corrected weight $\boldsymbol{F}^g$ is instantaneously balanced by hydrodynamic drag $\boldsymbol{F}^h_\alpha$ for each disc.
The simplest approximation for the hydrodynamic interaction experienced by a given disc would be to  sum the flows due to the other discs, treating each disc as a force monopole placed at its centroid. The resulting velocity, vorticity and rate of strain at the location of the disc in question translate, rotate and align it, leading to the equations of motion
\begin{equation}
    \label{eq:ValphaPt}
    \boldsymbol V_\alpha^{\text{pt}} = \frac{\boldsymbol{F}^g}{6\pi\mu a}\boldsymbol{\cdot}\left[\frac{\boldsymbol{p}_\alpha \boldsymbol{p}_\alpha}{X^A} + 
    \frac{(\boldsymbol{\delta} - \boldsymbol{p}_\alpha \boldsymbol{p}_\alpha)}{Y^A}\right]
    +
    \sum_{\beta \neq \alpha}^{N_d} \frac{\boldsymbol F^g}{8\pi\mu}\boldsymbol{\cdot}\boldsymbol{\mathsfbi{G}}(\boldsymbol x_{\alpha \beta})
    ,
\end{equation}
\begin{multline}
    \label{eq:dDotPt}
    \frac{d}{dt}\boldsymbol p_\alpha = \sum_{\beta\neq \alpha}^{N_d} \left(  \frac{\boldsymbol F^g \times \boldsymbol x_{\alpha \beta}}{8\pi\mu |\boldsymbol x_{\alpha \beta}|^3} \right) \times \boldsymbol p_\alpha 
    \\ 
    + \frac{3e^2}{2-e^2} \sum_{\beta\neq \alpha}^{N_d}\left(\frac{\boldsymbol F^g \boldsymbol{\cdot} \boldsymbol x_{\alpha \beta}}{8\pi\mu} \right) \left(\frac{\boldsymbol p_\alpha\boldsymbol{\cdot}\boldsymbol x_{\alpha \beta}}{|\boldsymbol x_{\alpha \beta}|^5} \right) [\boldsymbol x_{\alpha \beta} - \boldsymbol p_\alpha (\boldsymbol p_\alpha\boldsymbol{\cdot} \boldsymbol x_{\alpha \beta})], 
\end{multline}
where 
\begin{equation}
    \label{eq:oseenTensor}
    \boldsymbol{\mathsfbi{G}}(\boldsymbol x) = \frac{\boldsymbol \delta}{|\boldsymbol x|} + \frac{\boldsymbol x \boldsymbol x}{|\boldsymbol x|^3} 
\end{equation}
is the Green's function of the steady Stokes equation.
In \eqref{eq:ValphaPt} and \eqref{eq:dDotPt}, $\boldsymbol x_{\alpha \beta} \equiv \boldsymbol x_\alpha - \boldsymbol x_\beta$, 
$\mu$ is the dynamic viscosity of the fluid, $e = \sqrt{1-b^2/a^2}$ is the eccentricity, and 
\begin{gather}
    \label{eq:XaYa}
    X^A \equiv \frac{4}{3}e^3[(2e^2-1)K+e\sqrt{1-e^2}]^{-1},
    \\ Y^A \equiv \frac{8}{3}e^3[(2e^2+1)K-e\sqrt{1-e^2}]^{-1}, 
    \\{\rm with} \quad K \equiv \cot^{-1} \frac{\sqrt{1-e^2}}{e},    
\end{gather}
are resistance functions for an oblate spheroid \citep{kim2013microhydrodynamics}, which approximates the experimental discs with an aspect ratio of $b/a = 0.125$, corresponding to $e = 0.992$. 
The first term in \eqref{eq:ValphaPt} is the sedimentation velocity of an isolated disc falling under gravity and the second is the sum of disturbance fields produced by other discs, in the point-force approximation. The two terms in equation \eqref{eq:dDotPt} respectively are the contribution from the vorticity and the strain rate of the disturbance field on the orientational dynamics of the disc.

The point-force approximation captures the essential features of the periodic and unbounded dynamics of a pair of discs, and sharp transitions between these states, as the initial condition is varied \citep{chajwa2019kepler}. It also provides a basic framework to understand the linear instability and wave-like modes observed in a one-dimensional array of sedimenting discs \citep{chajwa2020waves}.
However, at late times, the initial lattice state is disrupted, giving rise to the formation of  configurations with small inter-particle separations, for which the point-force description is expected to be inadequate, 
since this approximation is accurate only up to $O(a/r)$ where $r=|\boldsymbol x_{\alpha \beta}|$. Therefore, for nearly touching disc configurations observed during late stages of instability in our experiments, 
we need to account for finite size effects, and to satisfy the no-slip boundary conditions on each particle. We use the method of reflections to study the close-range dynamics. This perturbative method is needed to iteratively satisfy the boundary conditions on the surface of each disc, with each iteration progressively correcting the errors from the previous one \citep{kim2013microhydrodynamics, kim1985sedimentation}. At the $n^{th}$ reflection, the velocity at the location of particle $\alpha$ satisfies the no-slip condition on it due to the velocities induced by all other particles calculated calculated at the $(n-1)^{th}$ reflection, and this corrected velocity of particle $\alpha$ will induce a small slip at the surfaces of all the other particles. As $n$ increases, the error goes to higher and higher orders in the interparticle spacing. We find that the first reflection, combined with a simple lubrication correction and near-contact repulsion, is sufficient to predict
the structural arrangement of discs observed in the experiments. The first reflection is accurate up to $O(a^3/r^3)$ in linear velocity and $O(a^4/r^4)$ in the angular velocity. 
The linear velocity $\boldsymbol{V}_\alpha$ and angular velocity $\boldsymbol\Omega_\alpha$ of a disc $\alpha$ can be expressed in terms of the hydrodynamic force $\boldsymbol{F}^h_\alpha$ and hydrodynamic torque $\boldsymbol{T}^h_\alpha$ acting on it, using the Faxén laws as \citep{kim2013microhydrodynamics}:
\begin{equation}
    \label{eq:Vfaxen}
    \boldsymbol{V}_\alpha = \frac{-\boldsymbol{F}^h_\alpha}{6\pi\mu a}\boldsymbol{\cdot}\left[\frac{\boldsymbol{p}_\alpha\boldsymbol{p}_\alpha}{X^A} + \frac{(\boldsymbol{\delta} - \boldsymbol{p}_\alpha\boldsymbol{p}_\alpha)}{Y^A}\right] \\
    + \frac{1}{2c}\int_{-c}^{c} d\xi \left\{1+ \frac{(c^2-\xi^2)}{4e^2}\nabla^2 \right\} \boldsymbol{u}^\infty_\alpha\bigg|_{\boldsymbol x_\alpha(\xi)},
\end{equation}
and 
\begin{multline}
    \label{eq:OmegaFaxen}
    \boldsymbol{\Omega}_\alpha = 
    \frac{-\boldsymbol{T}^h_\alpha}{8\pi\mu a^3}\boldsymbol{\cdot}\left[\frac{\boldsymbol{p}_\alpha\boldsymbol{p}_\alpha}{X^C} + \frac{(\boldsymbol{\delta} - \boldsymbol{p}_\alpha\boldsymbol{p}_\alpha)}{Y^C}\right] + 
    \frac{3}{8c^3}\int_{-c}^{c} d\xi (c^2-\xi^2) \boldsymbol{\nabla} \times \boldsymbol{u}^\infty_\alpha\bigg|_{\boldsymbol x_\alpha(\xi)} \\
    - \frac{3}{4c^3} \frac{e^2}{2-e^2} \int_{-c}^{c} d\xi \left\{ (c^2-\xi^2) \left[ 1 + (c^2-\xi^2)\frac{1}{8e^2}\nabla^2  \right] \right\}  \boldsymbol{p}_\alpha\times[\boldsymbol{E}^\infty_\alpha\bigg|_{\boldsymbol x_\alpha(\xi)}\boldsymbol{\cdot} \boldsymbol{p}_\alpha].
\end{multline}
Here $c \equiv a e$, $\boldsymbol{E}^\infty_\alpha$ is the strain rate associated with $u^\infty_\alpha$ and
\begin{gather}
    \label{eq:XcYc}
    X^C \equiv \frac{2}{3}e^3[K - e\sqrt{1-e^2}]^{-1},
    \\ Y^C \equiv \frac{2}{3}e^3(2-e^2)[e\sqrt{1-e^2}-(1-2e^2)K]^{-1}.
\end{gather}
In writing equations \eqref{eq:Vfaxen} and \eqref{eq:OmegaFaxen}, we have used the fact that the singularity distribution of an oblate spheroid (disc) can be written in terms of a line distribution placed along an imaginary focal length \citep{shatz2004singularity}. This requires us to evaluate the background incidence field on disc $\alpha$, $\boldsymbol{u}^\infty_\alpha$, at $\boldsymbol{x}_\alpha(\xi) \equiv \boldsymbol{x}_\alpha + i \xi \boldsymbol{p}_\alpha$.
Up to the first reflection correction, the background incidence field on disc $\alpha$, $\boldsymbol{u}_\alpha^\infty$, is given by the superposition of the disturbance fields generated by all the other discs, treating them as isolated and satisfying the no-slip boundary condition on their respective surfaces. Thus, we have
\begin{multline}
    \label{eq:uInf}
    \boldsymbol{u}_\alpha^\infty(\bx) \approx  \sum_{\beta \neq \alpha}^{N_d} \Bigg[\frac{-\boldsymbol{F}^h_\beta}{8\pi \mu} \boldsymbol{\cdot} \frac{1}{2c}\int_{-c}^{c} \left\{ 1 + \frac{(c^2-\xi^2)}{4e^2}\nabla^2 \right\} 
    \boldsymbol{\mathsfbi{G}}(\bx - \bx_\beta  - i\xi \bp_\beta)\, d\xi \\
    +\frac{1}{2}\frac{\boldsymbol{T}^h_\beta\times \boldsymbol\nabla}{8\pi \mu}\boldsymbol{\cdot} \frac{3}{4c^3}\int_{-c}^c (c^2-\xi^2)  \boldsymbol{\mathsfbi{G}}(\bx - \bx_\beta - i\xi\bp_\beta)\, d\xi
     \\
    - \boldsymbol{\mathsfbi{S}}_\beta^h\boldsymbol{\cdot}\boldsymbol{\nabla}\boldsymbol{\cdot} \frac{3}{4c^3}\int_c^c(c^2-\xi^2)\left\{1+\frac{c^2-\xi^2}{8e^2}\nabla^2 \right\} \boldsymbol{\mathsfbi{G}}(\bx - \bx_\beta - i\xi \bp_\beta)\, d\xi\Bigg],  
\end{multline}
where $N_d$ is the total number of discs and
\begin{equation}
    \label{eq:Sdef}
    \boldsymbol{\mathsfbi{S}}_\beta^h \equiv \frac{1}{2}\left(\frac{e^2}{2-e^2}\right)[(\boldsymbol{T}_\beta^h\times\bp_\beta)\bp_\beta + \bp_\beta(\boldsymbol{T}_\beta^h\times\bp_\beta)].
\end{equation}
Substituting equation \eqref{eq:uInf} into the equations \eqref{eq:Vfaxen} and \eqref{eq:OmegaFaxen}, we get the far field mobility matrix $\boldsymbol{\mathcal{M}}^\infty$ which relates the generalized velocities $\boldsymbol{\mathcal{V}}$ to generalized forces $\boldsymbol{\mathcal{F}}^h$ as:
\begin{equation}
    \label{eq:mobilityEqn}
    \boldsymbol{\mathcal{V}} \approx -\boldsymbol{\mathcal{M}}^\infty\boldsymbol{\cdot} \boldsymbol{\mathcal{F}}^h; \, \, \boldsymbol{\mathcal{V}} \equiv [\boldsymbol{V}_1 ...\, \boldsymbol{V}_{N_d}\, \boldsymbol{\Omega}_1 ...\, \boldsymbol{\Omega}_{N_d}]^T, \quad \boldsymbol{\mathcal{F}}^h \equiv [\boldsymbol{F}^h_1 ...\,  \boldsymbol{F}^h_{N_d}\, \boldsymbol{T}^h_1 ...\, \boldsymbol{T}^h_{N_d}]^T
\end{equation}
As the discs approach each other the far field mobility matrix has to be supplemented with lubrication effects which are dominant in the near contact configurations of the discs. The lubrication effects are preserved in the resistance formulation which can be incorporated by inverting $\boldsymbol{\mathcal{M}}^\infty$ and adding the near-contact resistance matrix $\boldsymbol{\mathcal{R}}^L$ \citep{durlofsky1987dynamic, claeys1993suspensions}. 
We use a simple model to compute the localized lubrication interaction between the nearly touching discs \citep{berne1972gaussian, gunther2014timescales, ladd2001lattice}. The hydrodynamic lubrication force on disc $\alpha$, nearly touching another disc $\beta$, is given by 
\begin{equation}
    \label{eq:Flub}
    \boldsymbol{F}^L_{\alpha\beta} = -\frac{6\pi\mu a [(\boldsymbol{V}_\alpha - \boldsymbol{V}_\beta)\boldsymbol{\cdot} \hat{\boldsymbol{\epsilon}}_{\alpha\beta}]}{8\sqrt{1-(\Gamma\bp_\alpha\boldsymbol{\cdot}\bp_\beta)^2}}\left(\frac{a}{|\boldsymbol{\epsilon}_{\alpha\beta}|} - \frac{1}{\Delta_c} \right)\hat{\boldsymbol{\epsilon}}_{\alpha\beta}, \quad \Gamma \equiv \frac{a^2-b^2}{a^2+b^2}, 
\end{equation}
where $\Delta_c = 2/3$ is the critical value below which the lubrication forces \eqref{eq:Flub} are computed \citep{gunther2014timescales, ladd2001lattice}. Here $\hat{\boldsymbol{\epsilon}}_{\alpha\beta} = \boldsymbol{\epsilon}_{\alpha\beta}/|\boldsymbol{\epsilon}_{\alpha\beta}|$ and $\boldsymbol{\epsilon}_{\alpha\beta}$ is the minimum separation vector between disc $\alpha$ and $\beta$ which can be computed by a simple extension of the method by \citep{claeys1993suspensions}, explained in the appendix.
The near-contact resistance matrix $\boldsymbol{\mathcal{R}}^L$ is computed by summing over the lubrication forces between the discs for which the minimum separation is less than $\Delta_c$.
Therefore, the mobility matrix $\boldsymbol{\mathcal{M}}$ constructed by taking into account the far field hydrodynamic interactions and near contact lubrication effects is given by
\begin{equation}
    \label{eq:Meqn}
    \boldsymbol{\mathcal{V}} \approx \boldsymbol{\mathcal{M}}\boldsymbol{\cdot} \boldsymbol{\mathcal{F}}, \quad
    \boldsymbol{\mathcal{M}} \equiv [(\boldsymbol{\mathcal{M}}^\infty)^{-1} + \boldsymbol{\mathcal{R}}^L]^{-1}.
\end{equation}
Here, $\boldsymbol{\mathcal{F}} = -\boldsymbol{\mathcal{F}}^h$ represents the generalized external force, which in our case consists solely of buoyancy-corrected weights, $\boldsymbol{F}^g$, for each disc, with zero torques.
Although the lubrication force mitigates violent particle approaches, numerical time-stepping can still cause unphysical particle overlap, which is generally prevented by incorporating repulsive near-contact forces \citep{butler2002dynamic, mari2014shear, drazer2002deterministic, gunther2014timescales, townsend2017mechanics}. We add a repulsive near-contact force that is supplemented with the buoyancy corrected weight $\boldsymbol{F}^g$ of each disc in obtaining the generalized external force $\boldsymbol{\mathcal{F}}$, given by \citep{mari2014shear, townsend2017mechanics}
\begin{equation}
    \label{eq:Frep}
    \boldsymbol{F}^r_{\alpha\beta} = \Tilde{k} e^{-\tau |\boldsymbol{\epsilon}_{\alpha\beta}|}\hat{\boldsymbol{\epsilon}}_{\alpha\beta}. 
\end{equation}
Based on test runs we choose $\tau=80/a$ and $\Tilde{k}=50$ in the simulations. The choice of a large coefficient in $\tau$ ensures that the repulsive force is only active when two particles get extremely close, and, together with the choice of a large $\Tilde{k}$, is operational only to prevent overlap of particles.

\section{Validation of the dynamics}
\label{sec:valdtn}

\begin{figure}
\centering
\includegraphics[width=0.9\textwidth]{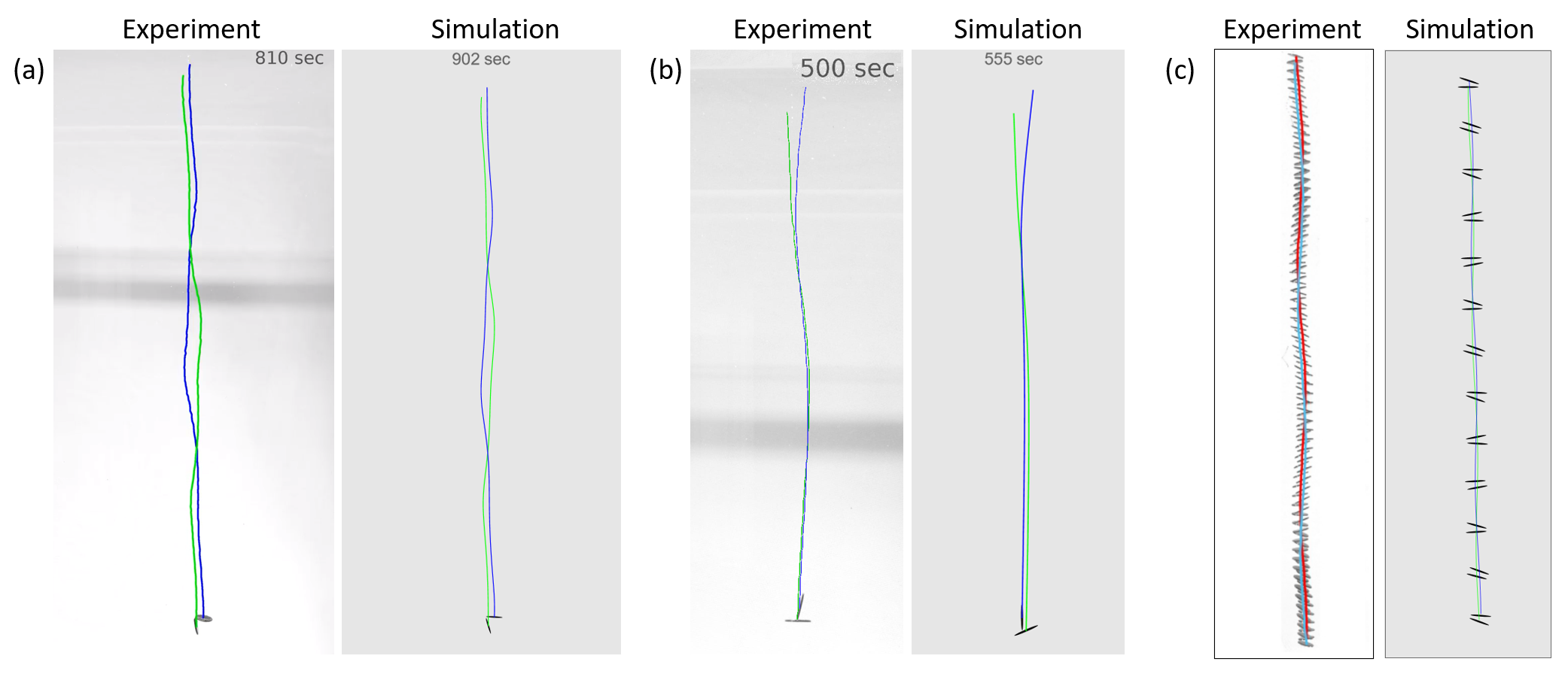}
\caption{\label{fig:compareExp} Comparison between experiments and simulations of pairs of discs sedimenting close to each other. (a) A bounded periodic oscillation of discs, categorized under `rocking dynamics' in \citet{chajwa2019kepler}. (b) `Hydrodynamic screening' where one falling disc enters the `hydrodynamic shadow' of the other, forming a `$\perp$' structure. This `$\perp$' configuration is often observed, both experimentally and through simulations, in a one-dimensional lattice of sedimenting discs. (c) Another case of `rocking dynamics', arising from the linear instability of an equilibrium configuration forming an `=' shape.}  
\end{figure}

The mobility matrix $\boldsymbol{\mathcal{M}}$ accounts for both far-field hydrodynamic interactions and near-contact lubrication effects, but omits
higher order effects coming from the second reflection and logarithmic corrections in the lubrication forces, which may not always be negligible. Nevertheless, the current approximation adequately captures the qualitative features of the cluster formation observed in our experiments. We therefore expect that higher-order corrections would primarily improve quantitative aspects, such as sedimentation velocities, but approaching such a precision is beyond the scope of this work. 
Our simulation model does not completely capture all the late-time dynamics of cluster evolution seen in our experiments, and we believe the differences are because of the finiteness of the container. Specifically, while pairs of discs that settle into a `$\perp$’ structure largely maintain this configuration thereafter in simulations, experimental observations show that only a few discs retain the `$\perp$’ arrangement for significantly long after pairing. However, in experiments involving isolated disc pairs, the long-term stability of the `$\perp$’ structure has been confirmed (see figure \ref{fig:compareExp})(b). In any case, our analysis focuses on cluster formation, and we refrain from drawing conclusions about their late-time evolution based on our simulations. 

Figure \ref{fig:compareExp}(a) demonstrates that our model successfully captures near-contact periodic oscillations, along with the phenomenon of `hydrodynamic screening' shown in Figure \ref{fig:compareExp}(b), where one disc enters the `hydrodynamic shadow' of another, forming a `$\perp$' structure. 
The initial configuration for figure \ref{fig:compareExp}(a) is: $\bx_{21}= \{0.68, 0, 1.18 \}$, $\bp_1 = \{-0.04, 0, 0.99\}$ and $\bp_2 = \{0.946, 0, 0.324\}$. Figure \ref{fig:compareExp}(b) has the initial configuration: $\bx_{21}= \{1.5, 0, 1.813 \}$, $\bp_1 = \{0.879, 0, 0.476\}$ and $\bp_2 = \{0.965, 0, -0.26\}$.
It may be noticed from the figure that quantitative details such as the time it takes for the discs to fall through the same distance in experiments and simulations differ. 
An interesting case is that of the `rocking dynamics' \citep{chajwa2019kepler} shown in Figure \ref{fig:compareExp}(c). 
This dynamics can be viewed as the evolution of perturbations from a steady-state configuration in which the discs are vertically stacked, with their symmetry axes aligned along gravity, forming an `=' shape. This enables a linear stability analysis of such a structural arrangement, which we shall perform in the following subsection, revealing that at least the first reflection is essential to capture the periodic oscillations characteristic of the rocking dynamics, a phenomenon that the point-force approximation (equations \eqref{eq:ValphaPt} and \eqref{eq:dDotPt}) fails to predict at small separations. 

\subsection{Stability analysis of the `=', or horizontal pair, configuration}
\label{sec:stabilityEqualShape}

The configuration where both discs are horizontal, and one  lies above the other at a vertical distance of $z^*$ times the size of the disc $a$, is studied in this subsection in some detail, because it is a canonical example of pair interactions, and because 
it analytically demonstrates the inability of the point force approximation to accurately capture close-range dynamics.
The rocking dynamics shown in figure \ref{fig:compareExp}(c) are periodic oscillations in the orientations and positions of the pair of discs about the `$=$' configuration represented by $(|\boldsymbol{p}_1\boldsymbol{\cdot} \boldsymbol{g}|,\, |\boldsymbol{p}_2\boldsymbol{\cdot} \boldsymbol{g}|)=(1,1)$. Here $\boldsymbol{g}$ is a unit vector along gravity (along the $-z$ direction). 
The `$=$' configuration is a fixed point in the frame falling with the pair of discs. 
The system is non-dimensionalized using the semi-major axis $a$ of the discs as the length scale and $\tau_p={6\pi\mu a^2}/F$ as the time scale, where $F$ is the buoyancy-corrected weight of the spheroids. 
We shall henceforth work with dimensionless quantities, unless stated otherwise.
Let the position of the disc 2 (upper disc) relative to disc 1 (lower disc) be given by $(\delta x,z^* + \delta z)$ and their corresponding orientation vectors be given by: $\boldsymbol{p}_1=(\cos(\pi/2+\delta\theta_1), \sin(\pi/2+\delta\theta_1))$ and  $\boldsymbol{p}_2=(\cos(\pi/2+\delta\theta_2), \sin(\pi/2+\delta\theta_2))$; with their fixed point position and orientations being $(0,z^*)$ and $\boldsymbol{p}_1^*=\boldsymbol{p}_2^*=(0,1)$, respectively. The dynamics of the pair, using the first reflection, leads to the following linear stability analysis:
\begin{equation}
    \label{eq:Pt=}
    \frac{d}{dt}\begin{pmatrix}
        \delta x \\
        \delta \theta_1 \\
        \delta \theta_2 
    \end{pmatrix} =
    \mathsfbi{B}_{R_1}
    \begin{pmatrix}
        \delta x \\
        \delta \theta_1 \\
        \delta \theta_2 
    \end{pmatrix};\quad 
    \mathsfbi{B}_{R_1}=\begin{pmatrix}
        0 & \kappa_0(e) & -\kappa_0(e) \\
        J_1 & J_2 & J_3 \\
        -J_1 & -J_3 & -J_2 
    \end{pmatrix},
\end{equation}
where 
\begin{equation}
    \label{eq:kappa0}
    \kappa_0(e) \equiv \left(\frac{1}{Y^A}-\frac{1}{X^A}\right),
\end{equation}
\begin{equation}
    \label{eq:J1}
    J_1 \equiv \frac{-3}{4} \int_{-1}^1\frac{3}{8}d\xi_1\int_{-1}^1 d\xi_2 \frac{1-\xi_1^2}{\Tilde\xi_0^3} \left\{ 1-\frac{e^2}{2-e^2} \left[ -3 +\frac{24\kappa_1}{\Tilde\xi_0^2} \right] \right\},
\end{equation}
\begin{multline}
    \label{eq:J2}
    J_2 \equiv \frac{-3}{4} \int_{-1}^1\frac{3}{8}d\xi_1\int_{-1}^1 d\xi_2 \frac{1-\xi_1^2}{\Tilde\xi_0^3} \Bigg\{ i e\xi_1 - \frac{e^2}{2-e^2} \Bigg[ (z^*+i e\xi_2)\left( -3 +\frac{30\kappa_1}{\Tilde\xi_0^2} \right) \\
    - \frac{6\kappa_1 (2z^*-ie\xi_1+2ie\xi_2)}{\Tilde\xi_0^2} \Bigg] \Bigg\},
\end{multline}
and
\begin{equation}
    \label{eq:J3}
    J_3 \equiv \frac{-3}{4} \int_{-1}^1\frac{3}{8}d\xi_1\int_{-1}^1 d\xi_2 \frac{1-\xi_1^2}{\Tilde\xi_0^3} \left\{-1+\frac{e^2}{2-e^2} \left[ -3 +\frac{24\kappa_1}{\Tilde\xi_0^2} \right] \right\}i e\xi_2.
\end{equation}
For shorthand notation we define the following quantities appearing in integrands:
\begin{equation}
    \label{eq:kappaParams}
\Tilde\xi_0\equiv \sqrt{(z^*+ie(\xi_2-\xi_1))^2} \quad {\rm and} \ \ \kappa_1 \equiv  \frac{3/2-\xi_1^2-\xi_2^2}{4}.
\end{equation}
The vertical perturbation $\delta z$ appears only at second order in the algebra, and therefore does not feature in the linear stability analysis. The corresponding Jacobian matrix of the perturbation evolution in the point force limit, obtained from equations \eqref{eq:ValphaPt} and \eqref{eq:dDotPt},  is given by
\begin{equation}
    \label{eq:Bpt}
    \mathsfbi{B}_{pt} = 
\begin{pmatrix}
    0 & \kappa_0(e) & -\kappa_0(e) \\[1ex]
    \frac{-3}{4z^3}\left[1+\frac{3e^2}{2-e^2}\right] & \frac{-3}{4z^2}\left[\frac{3e^2}{2-e^2}\right] & 0 \\[1ex]
    \frac{3}{4z^3}\left[1+\frac{3e^2}{2-e^2}\right] & 0 & \frac{3}{4z^2}\left[\frac{3e^2}{2-e^2}\right].
\end{pmatrix}
\end{equation}

\begin{figure}
\centering
\includegraphics[width=0.9\textwidth]{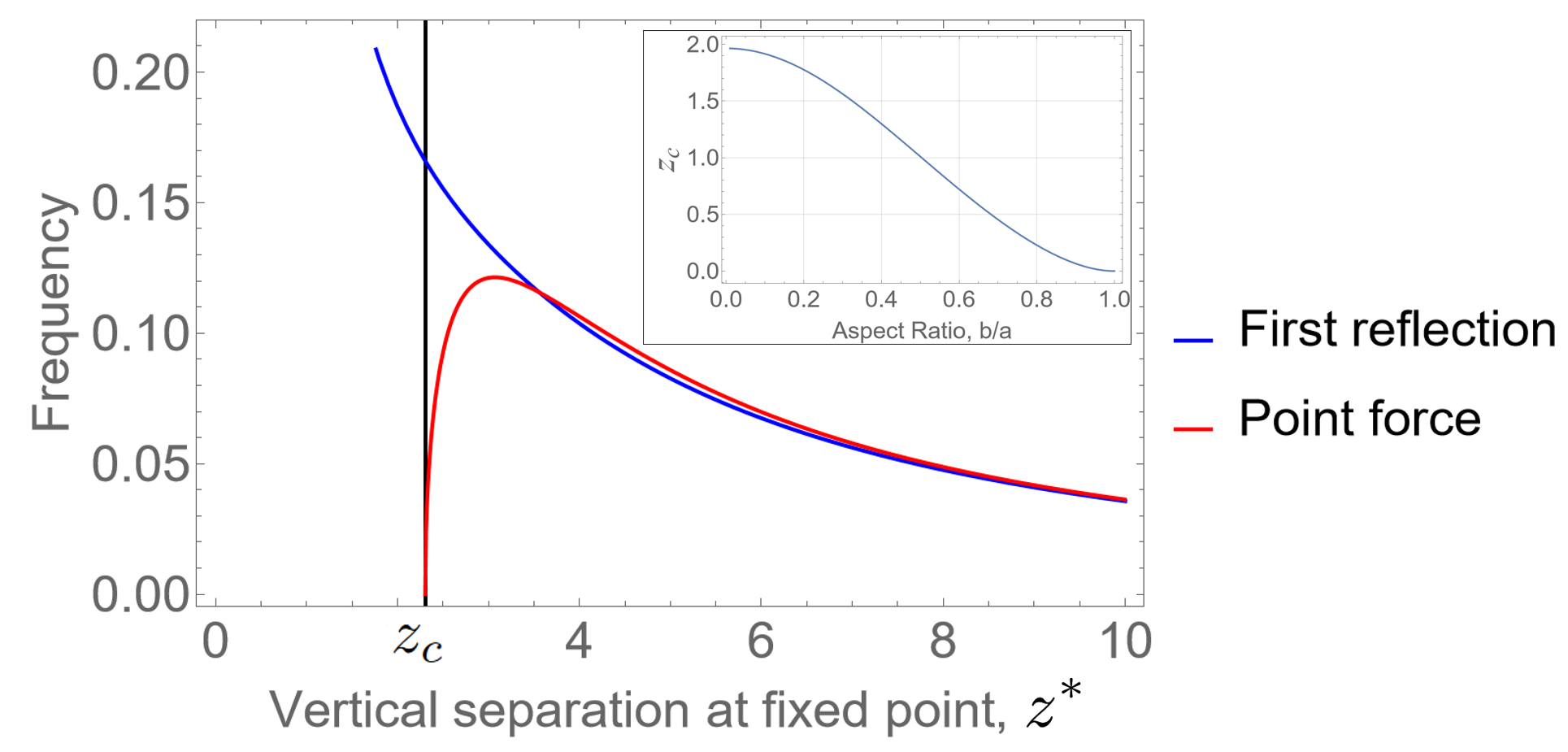}
\caption{\label{fig:rockOmega} The frequency of oscillations in the rocking dynamics about the vertically stacked, or `=' configuration, as a function of vertical separation $z^*$ at the fixed point. The aspect ratio here is  $b/a=0.125$, which is close to our experimental discs. The black vertical line denotes $z=z_c$ below which the point force approximation fails to predict the rocking dynamics. The inset shows the dependence of the $z_c$ on the aspect ratio.}
\end{figure}

Under the first reflection approximation, the eigenvalues of $\mathsfbi{B}_{R_1}$ are purely imaginary for any $z^*$ and hence the linear dynamics is periodic about the `$=$' configuration. The oscillation frequency of this rocking dynamics is shown in figure \ref{fig:rockOmega}. The point force approximation, on the other hand, leads us to the incorrect conclusion that the fixed point `$=$' is a saddle below a certain value of $z=z_c$, given by
\begin{equation}
    \label{eq:zc}
    z_c \equiv \left[\frac{16}{27}\left( \frac{2+e^2}{e^4} - 1\right)\kappa_0(e) \right]^{-1}.
\end{equation}
A linear stability analysis using the point force approximation thus cannot predict the rocking dynamics seen in experiment for small disc separations, which shows that the first reflection correct is crucial in this regime.
Note that $z^*$ is bounded below by $2b/a$ which corresponds to case where the discs just touch.

\section{Planar dynamics of a one-dimensional lattice of sedimenting discs}
In this section we study the sedimentation dynamics of a regular one-dimensional lattice of identical spheroids, with their symmetry axes aligned with the horizontal. The corresponding problem for spheres was studied by \citet{crowley1971viscosity}, who established that any arbitrary perturbation in the initial positions of the spheres will always grow exponentially, i.e., the system is always unstable. 
Crowley also proposed a mechanism to explain the growth of perturbations, which leads to the formation of clumps of spheres. \citet{chajwa2020waves} investigated the planar sedimentation dynamics of a regular one-dimensional lattice of identical spheroids, both experimentally as well as theoretically, and showed, in contrast to spheres, that a system of spheroids is not always exponentially unstable: there is a regime where linear perturbations display algebraic rather than exponential growth. In the algebraically growing regime, \citet{chajwa2020waves} found that the modal perturbations are neutrally stable. 

The analysis presented in \citet{chajwa2020waves} was based on the point force approximation (see equations \eqref{eq:ValphaPt} and \eqref{eq:dDotPt}) with nearest-neighbour interactions. 
In this section we show that this picture is crucially modified 
when we take into account more accurate hydrodynamic interactions through the first reflection. We find that the qualitative behaviour of modes changes from pure oscillations to weak exponential growth -- the sharp boundary between these two regions is lifted, and the growth rate changes smoothly over the entire regime 
rather than becoming strictly zero.

\begin{figure}
\centering
\includegraphics[width=0.9\textwidth]{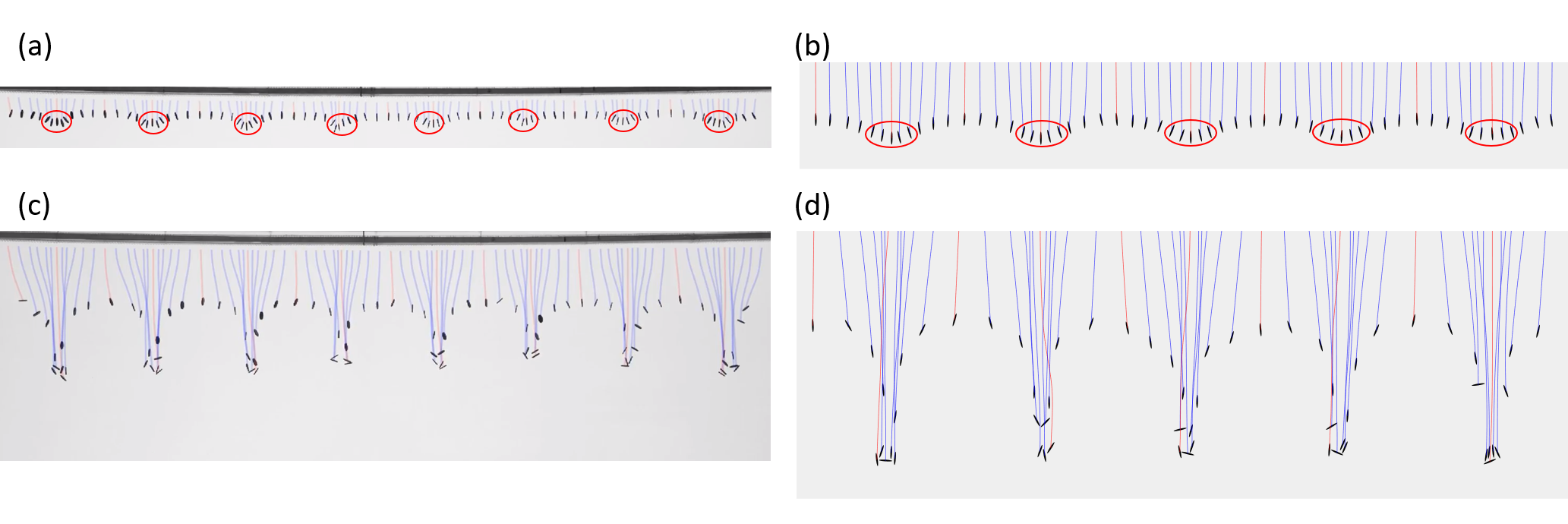}
\caption{\label{fig:expClump} 
Experiments [(a) and (c)], and simulations [(b) and (d)] showing clumping due to the Crowley mechanism. 
[(a) and (c)] are aligned vertically, as are [(b) and [(d)], to highlight that the locations where the clumps form correspond to the red ovals in (a) and (b).
The lattice spacing and the dominant perturbation wavenumber are $\Tilde{d}=1.875$ and $q=\pi/6$ respectively. (a) and (b) show early evolution ($t\sim 5$). The valleys, marked by red ovals, represent regions where the discs begin to clump together.
(c) and (d) show late time evolution ($t\sim 15$). Clumps, occurring at the valleys, consist of three or more discs.
}
\end{figure}

\begin{figure}
\centering
\includegraphics[width=0.9\textwidth]{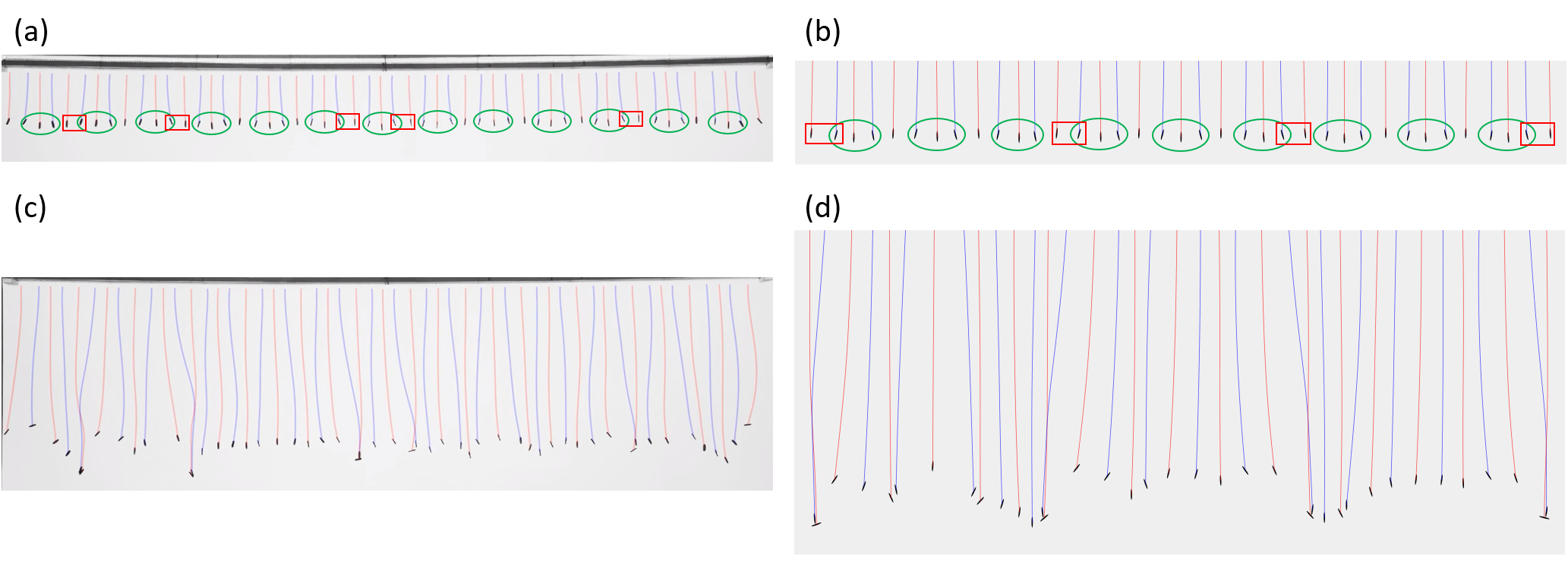}
\caption{\label{fig:lowQClump} 
Experiments [(a) and (c)], and simulations [(b) and (d)] showing pairing due to the drift mechanism. 
[(a) and (c)] are aligned vertically, as are [(b) and [(d)], to highlight that the locations where the pairs form correspond to the red boxes in (a) and (b).
The lattice spacing and the perturbation wavenumber are $\Tilde{d}=3.75$ and $q=\pi/2$ respectively. (a) and (b) show initial evolution ($t\sim 20$). 
The valleys are highlighted with green ovals, while the discs that ultimately pair up are enclosed within red rectangular boxes, and seen to be located away from the valleys.
(c) and (d) show late time evolution ($t\sim45$). There is pairing but no clumping, and pairing dynamics is characterized by two discs coming together in the form of a `$\perp$', or inverted `T'.
}
\end{figure}

For ease of distinguishing we refer to clusters of three or more spheroids as clumps, and to clusters of two spheroids as pairs. 
Visuals of disc clusters reveal that when clustering occurs, clumping takes place in the valleys when the Crowley mechanism dominates over the drift mechanism 
(figure \ref{fig:expClump}). However, when the drift mechanism is stronger than the Crowley mechanism, there are no clumps, but pairing occurs away from the valley. Pairs typically take on a `$\perp$' shape, see figure \ref{fig:lowQClump}. 

\begin{figure}
\centering
\includegraphics[width=0.6\textwidth]{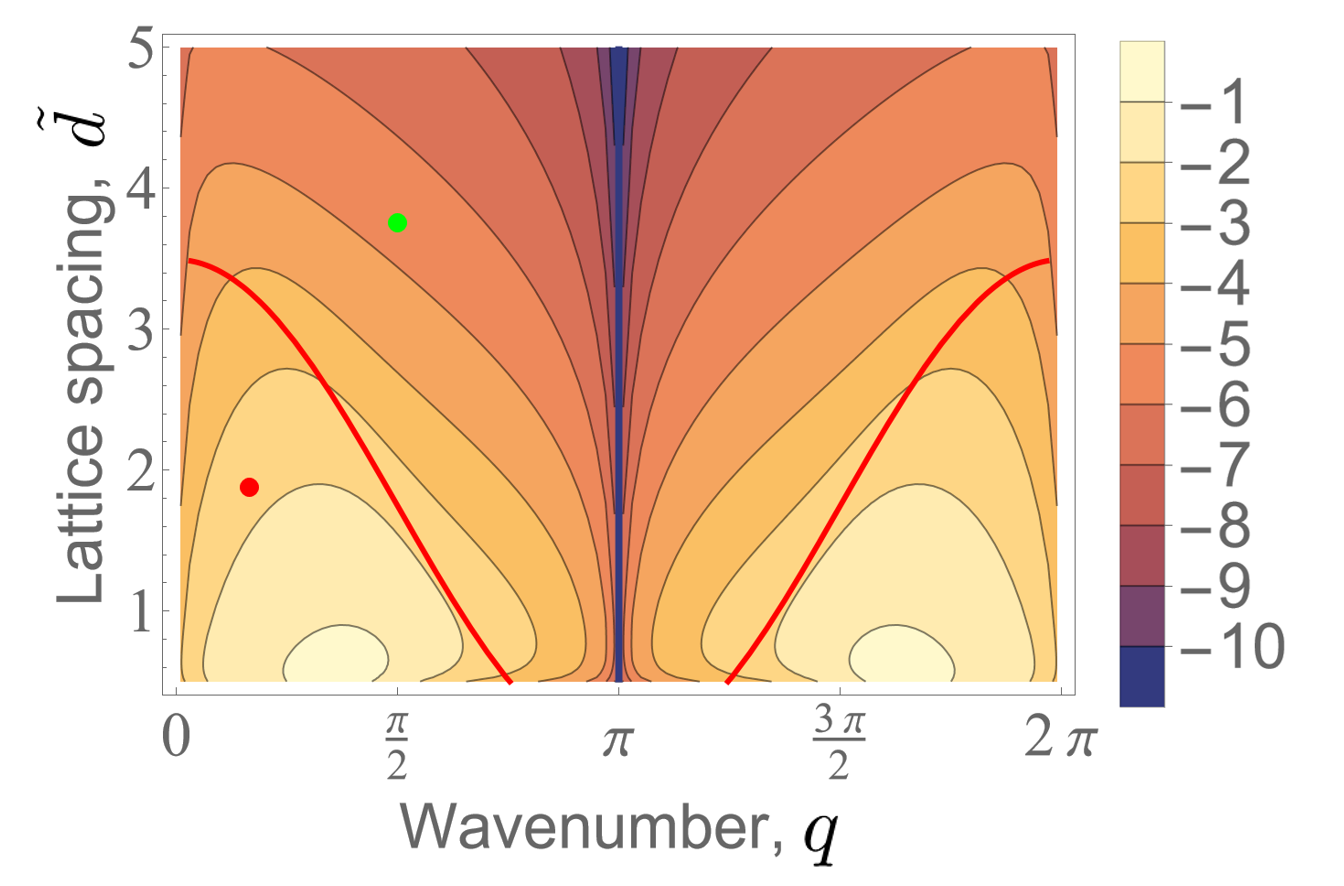}
\caption{\label{fig:logGwthRate} Contour plot of the log of the growth rate $\sigma$ (the only positive real part among the eigenvalues of $\mathsfbi{A}_{R_1}$) for aspect ratio $b/a = 0.125$. 
Note that $\sigma\to 0$ as $\Tilde{d}\to \infty$. Moreover, for $q=\pi$, we have $\sigma=0$ for all $\Tilde{d}$. The red curve denotes the critical lattice spacing $\Tilde{d}_c$ above which the point force approximation \citep{chajwa2020waves} predicts neutral stability. The nonlinear evolution corresponding to the red dot, where $(q,\Tilde{d})=(\pi/6,1.875)$ and the green dot, where $(q,\Tilde{d})=(\pi/2,3.75)$, were shown to support clustering and pairing respectively (figures \ref{fig:expClump} and \ref{fig:lowQClump}).}
\end{figure}

To numerically investigate the distinction between clumps and pairs, as observed in the experiments, we use two different lattice spacings and wavenumbers, as indicated by the two coloured dots in the figure \ref{fig:logGwthRate}. 
The simulations are conducted by solving equation \eqref{eq:Meqn} as a system of ordinary differential equations using the Runge–Kutta–Fehlberg method for time integration \citep{fehlberg1969low}. Adaptive time stepping is employed with an error tolerance of $10^{-4}$ and a maximum step size of $0.1$.
In the Crowley-mechanism-dominated regime with lattice spacing $\Tilde{d}=1.875$ and dominant perturbation wavenumber $q=\pi/6$ (red dot in figure \ref{fig:logGwthRate}), the resulting clumps formed at the valleys are shown in figure \ref{fig:expClump}. In the drift-mechanism-dominated regime with lattice spacing $\Tilde{d}=3.75$ and dominant perturbation wavenumber $q=\pi/2$ (green dot in figure \ref{fig:logGwthRate}), pairwise clumping is observed as shown in figure \ref{fig:lowQClump}.

There are three degrees of freedom per disc, the horizontal position $x_\alpha$ of disc $\alpha$, its vertical position $z_\alpha$, and its orientation angle $\theta_\alpha$ with respect to the horizontal ($x$) axis.
About the fixed point $\Tilde{d}(\mathbb{Z},0,0)$\footnote{This fixed point is in the frame of reference falling with the average terminal velocity of the undisturbed lattice.}, $\Tilde{d}$ being the dimensionless lattice spacing, an infinitesimal perturbation of the disc $\alpha\in\mathbb{Z}$ is given by $(\hat{x}_q (t),\hat{z}_q (t),\hat{\theta}_q (t)) e^{iq\alpha}$, where $q$ is the dimensionless perturbation wavenumber. Under the first reflection and nearest-neighbour interactions, this perturbation evolves as
\begin{equation}
    \label{eq:XDotLin}
    \frac{d}{dt}\begin{bmatrix}
        \hat{x}_q \\[6pt]
        \hat{z}_q \\[6pt]
        \hat{\theta}_q \\[6pt]
    \end{bmatrix}
   = \mathsfbi{A}_{R_1}
    \begin{bmatrix}
        \hat{x}_q \\[6pt]
        \hat{z}_q \\[6pt]
        \hat{\theta}_q \\[6pt]
    \end{bmatrix}; \quad
    \mathsfbi{A}_{R_1} \equiv \begin{bmatrix}
        0 & I_{12} & \kappa_0(e) + I_{13} \\[6pt]
        I_{21} & 0 & 0 \\[6pt]
        I_{31} & 0 & 0
    \end{bmatrix},
\end{equation}
where $\kappa_0$ is given by equation \eqref{eq:kappa0}.
The terms in the matrix $\mathsfbi{A}_{R_1}$ are given by
\begin{equation}
    \label{eq:I12}
    I_{12} \equiv \dfrac{3}{2} \int_{-1}^1\frac{1}{2}d\xi_1\int_{-1}^1\frac{1}{2}d\xi_2 i\sin q \frac{i e \xi_{12}-\Tilde{d}}{\xi_{0}^3}\left[1-\frac{6\kappa_2}{\xi_0^2} \right], 
\end{equation}

\begin{equation}
    \label{eq:I21}
    I_{21} \equiv -\dfrac{3}{2} \int_{-1}^1\frac{1}{2}d\xi_1\int_{-1}^1\frac{1}{2} d\xi_2 i\sin q \frac{i e \xi_{12}-\Tilde{d}}{\xi_{0}^3}\left[1+\frac{6\kappa_2}{\xi_0^2} \right], 
\end{equation}

\begin{equation}
    \label{eq:I13}
    I_{13} \equiv -\dfrac{3}{2} \int_{-1}^1\frac{1}{2}d\xi_1\int_{-1}^1 \frac{1}{2}d\xi_2 i e[\xi_1-\xi_2\cos(q)] \frac{i e \xi_{12}-\Tilde{d}}{\xi_{0}^3}\left[1-\frac{6\kappa_2}{\xi_0^2} \right], 
\end{equation}

\begin{multline}
    \label{eq:I31}
    I_{31} \equiv \frac{3}{2} \int_{-1}^1\frac{3}{8}d\xi_1\int_{-1}^1 d\xi_2 \frac{1-\xi_1^2}{\xi_0^3} \Bigg\{[1-\cos(q)]\left(-2+\frac{6e^2
    \kappa_1}{(2-e^2)\xi_0^2} \right) + \\
    \frac{30e^2\kappa_1}{(2-e^2)\xi_0^2} \frac{i\sin q}{(i e\xi_{12}-\Tilde{d})} \Bigg\}, 
\end{multline}
where $\xi_{12} \equiv \xi_1-\xi_2$, $\xi_0\equiv \sqrt{(i e\xi_{12}-\Tilde{d})^2}$ and $\kappa_2 \equiv (2-\xi_1^2-\xi_2^2)/4$.
The eigenvalues of $\mathsfbi{A}_{R_1}$ are $\{0, \sigma+i\omega, -\sigma-i\omega \}$ where $\sigma$ represents the growth rate of perturbations, and $\omega$ denotes the oscillation frequency. The structure of the eigenvalues ensures that the system is always unstable provided $\sigma \neq 0$.

\begin{figure}
\centering
\includegraphics[width=0.62\textwidth]{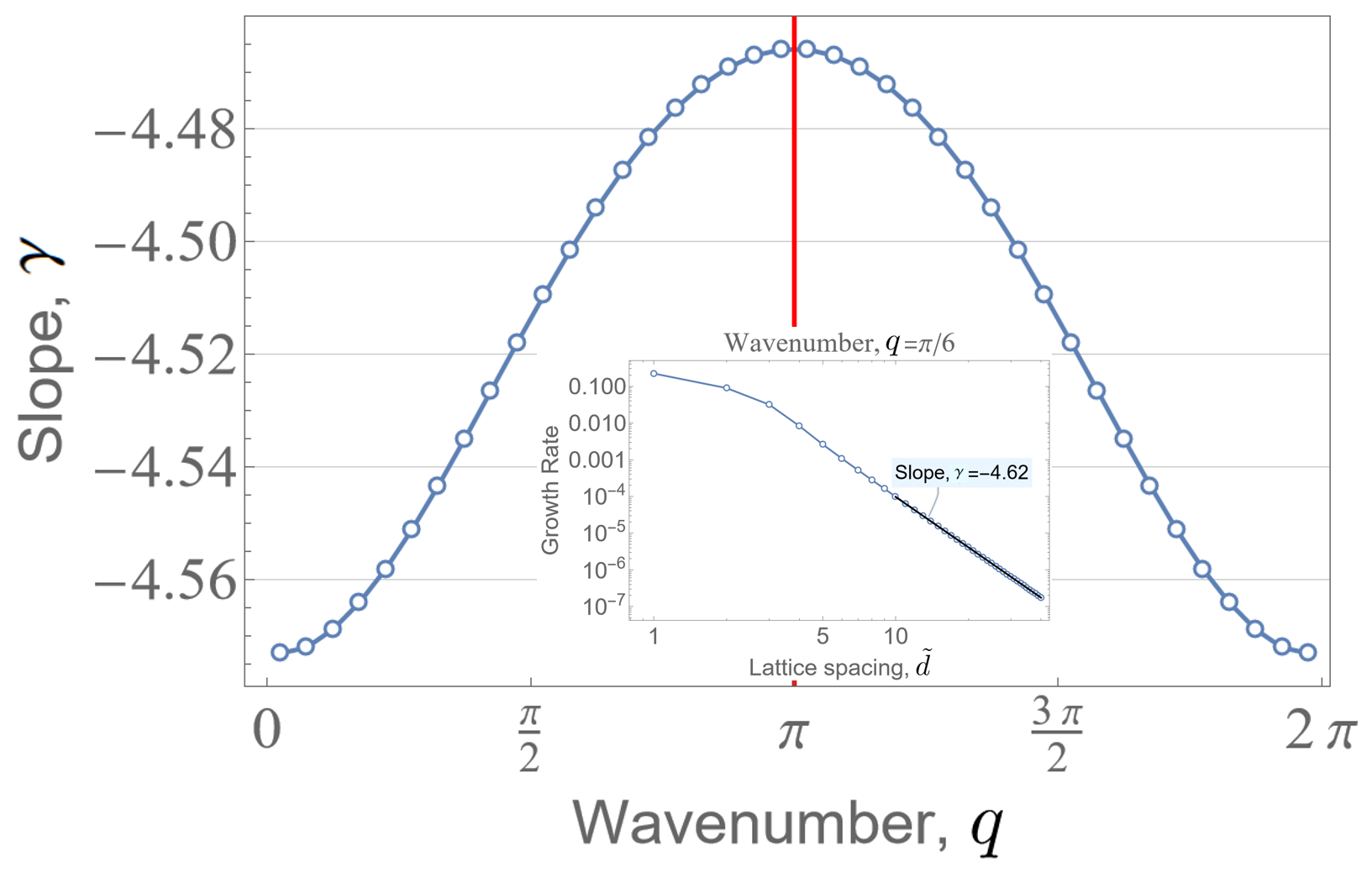}
\caption{\label{fig:decayWithD} Inset: perturbation growth rate $\sigma$ as a function of the lattice spacing for a wavenumber $q=\pi/6$ and aspect ratio $b/a=0.125$. For large lattice spacing $\Tilde{d}$, the growth rate shows a power law behaviour $\sigma \sim \Tilde{d}^\gamma$ with $\gamma\approx -4.5$. The main plot shows variation of the exponent $\gamma$ with wavenumber $q$. $\gamma$ is not defined at $q=\pi$ (indicated by the red line), since $\sigma=0$ at $q=\pi$.  Note that for a lattice of spheres the growth rate decays much more slowly, as $\sigma \sim \Tilde{d}^{-2}$ for all wavenumbers. 
}
\end{figure}

On the other hand, under the point-force approximation, the  corresponding Jacobian matrix $\mathsfbi{A}_{pt}$ for the evolution of perturbations is given by \citep{chajwa2020waves} 
\begin{equation}
    \label{eq:Apt}
    \mathsfbi{A}_{pt} \equiv \begin{bmatrix}
        0 & \frac{-3 i \sin q}{2\Tilde{d}^2} & \kappa_0(e) \\[6pt]
         \frac{3 i \sin q}{2\Tilde{d}^2} & 0 & 0 \\[6pt]
         \frac{-6\sin^2 q/2}{\Tilde{d}^3} & 0 & 0
        \end{bmatrix}.
\end{equation}
For a given wavelength, the point-force approximation predicts a neutrally stable regime characterized by a zero growth rate beyond a critical lattice spacing $\Tilde{d}_c$, given by
\begin{equation}
    \label{eq:dCrit}
    \Tilde{d}_c = \frac{3}{8\kappa_0(e)} \frac{\sin^2 q}{\sin^2 q/2}.
\end{equation} 
Figure \ref{fig:logGwthRate} shows the logarithm of the growth rate of perturbations as a function of lattice spacing $\Tilde{d}$ and perturbation wavenumber $q$. While the point-force approximation predicts zero growth rate above the red curve, the first reflection has small but finite growth rate at high lattice spacings. The growth rate shows a power law behaviour $d^{-\gamma}$ for large lattice spacing with the dependence of the power on the wavenumber shown in figure \ref{fig:decayWithD}. 
Numerical data show that the scaling exponent in the large-$d$ limit is best fit by $\gamma \approx -4.5$ across perturbation wavenumbers, for a large range of spheroids with aspect ratio ranging from 0.19 to 0.8.
\begin{figure}
\centering
\includegraphics[width=0.8\textwidth]{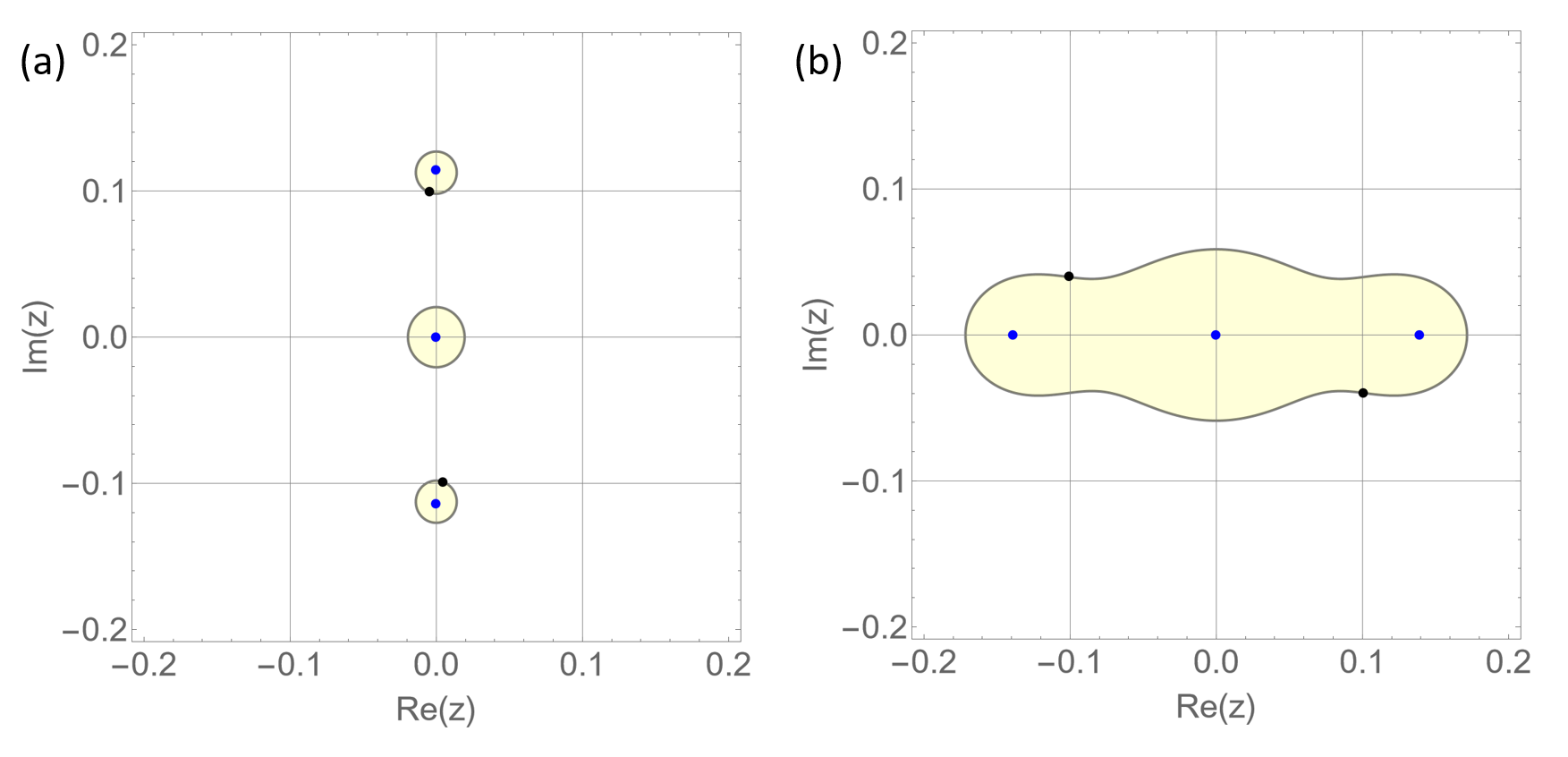}
\caption{\label{fig:pseudoSpectra} $\varepsilon$-pseudospectrum $\sigma_{\varepsilon}(\mathsfbi{A}_{pt})$ of $\mathsfbi{A}_{pt}$ for the aspect ratio $b/a = 0.125$. The blue dots denote the eigenvalues of $\mathsfbi{A}_{pt}$ and the black dots denote the eigenvalues of $\mathsfbi{A}_{R_1}$. The matrix perturbation size $\varepsilon$ leads to spreading of eigenvalues inside the yellow regions. 
The values of $\varepsilon$ are chosen so that the black dots lie at the boundary of the yellow regions, indicating that the first reflection correction to $\mathsfbi{A}_{pt}$ are of the order of $\varepsilon$.
a) $\varepsilon=4.8\times10^{-3}$ 
for the set of parameters $(q,\Tilde{d})=(\pi/2,3.75)$ (green dot in figure \ref{fig:logGwthRate}) 
b) $\varepsilon=1.2\times10^{-2}$ for the set of parameters $(q,\Tilde{d})=(\pi/6,1.875)$ (red dot in figure \ref{fig:logGwthRate}). Note that the pseudospectra are much larger than $\epsilon$.
}
\end{figure}

The first reflection thus shows a qualitative change in the stability of the system as compared to the point-force approximation: the system is now linearly unstable at any lattice spacing, although with 
a growth rate that approaches zero with increasing spacing. 
An examination of the point-force approximation shows that this is to be expected. The non-normality of $\mathsfbi{A}_{pt}$ provides for possible large change in eigenvalues for a small perturbation in the stability operator ($\mathsfbi{A}_{pt}$ becoming $\mathsfbi{A}_{R_1}$).
Since in the neutral regime the eigenvalues of $\mathsfbi{A}_{pt}$ lie on the imaginary axis, its pseudospectrum must protrude into the positive real axis, and $\mathsfbi{A}_{R_1}$ is a member of the pseudospectrum of $\mathsfbi{A}_{pt}$ which produces instability.  The $\varepsilon$-pseudospectrum $\sigma_{\varepsilon}(\mathsfbi{A}_{pt})$ of $\mathsfbi{A}_{pt}$ is defined as  \citep{trefethen1997pseudospectra}:
\begin{equation}
    \label{eq:pseudoSpectra}
    \sigma_{\varepsilon}(\mathsfbi{A}_{pt}) \equiv \{z \in \mathbb{C}\, :\, ||(z-\mathsfbi{A}_{pt})^{-1}||>\varepsilon^{-1} \}.
\end{equation}

Figure \ref{fig:pseudoSpectra}(a) and (b) show the pseudospectra of $\mathsfbi{A}_{pt}$ corresponding to the conditions of figures \ref{fig:expClump} and \ref{fig:lowQClump} respectively. In each case $\varepsilon$ is chosen such that the largest eigenvalue of $\mathsfbi{A}_{R_1}$ lies on the boundary of the pseudospectrum.  
While the oscillation frequency remains close to its original value, a non-zero growth rate appears which changes the dynamics qualitatively. Therefore, in the drift regime (green dot in figure \ref{fig:logGwthRate}), the first reflection correction has a gentle de-stabilizing effect. In contrast to this, the first reflection has a significant stabilizing effect in the Crowley regime, and lies on a larger $\varepsilon$-pseudospectrum in figure \ref{fig:pseudoSpectra}(b). The net growth rate remains positive (red dot in figure \ref{fig:logGwthRate}). 

\begin{figure}
\centering
\includegraphics[width=0.9\textwidth]{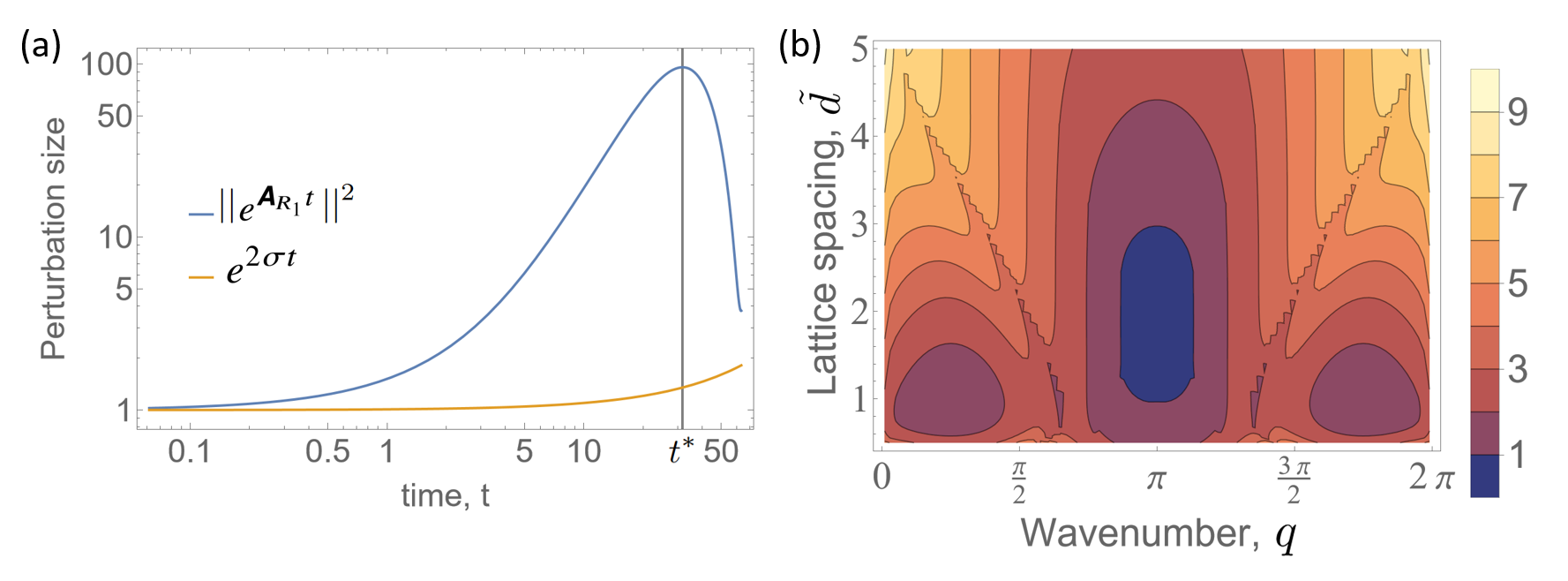}
\caption{\label{fig:GvsExp} (a) Time evolution of the matrix norm of $e^{\mathsfbi{A}_{R_1}t}$ and the exponential growth $e^{2\sigma t}$ for the aspect ratio $b/a=0.125$. The non-normal nature of $\mathsfbi{A}_{R_1}$ leads to transient growth of non-modal perturbations which can be much larger than the exponential growth. The parameters $(q,\Tilde{d})$ used to calculate $\mathsfbi{A}_{R_1}$ correspond to the green dot in figure \ref{fig:logGwthRate}. The non-modal growth reaches it maximum at time $t=t^*$. (b) Contour plot of the log of $G_r$ defined in equation \eqref{eq:Growthratio}. In regimes of small $\log(G_r)$, the growth is exponential but very slow. Large values of $\log(G_r)$ indicate that transient algebraic growth will dominate.
}
\end{figure}

If only eigenvalue-based growth were in play, the extremely slow growth rate for large lattice spacing would contribute negligibly to perturbation growth during the time of our simulations. However, a large algebraic transient growth of the perturbations happens in this regime, due to the non-normal nature of $\mathsfbi{A}_{R_1}$. To highlight this, we look at the matrix norm of $e^{\mathsfbi{A}_{R_1}t}$ for time $t\in [0,2\pi/\omega]$, where $\omega$ is the oscillation frequency given by the imaginary parts of non-zero eigenvalues of $\mathsfbi{A}_{R_1}$, see figure \ref{fig:GvsExp}. The maximum non-modal growth can be quantified by the quantity $G_r$ defined as:
\begin{equation}
    \label{eq:Growthratio}
    G_r \equiv \frac{ ||e^{\mathsfbi{A}_{R_1} t^*} ||^2}{e^{2\sigma t^*}}; \quad t^* \equiv \argmax_{t\in[0,2\pi/\omega]}||e^{\mathsfbi{A}_{R_1}t}|| ,
\end{equation}
where $\sigma$ is the growth rate given by real parts of eigenvalues of $\mathsfbi{A}_{R_1}$ and the matrix norm can be obtained using the singular value decomposition. Figure \ref{fig:GvsExp} shows the logarithm of $G_r$ as a function of lattice spacing $\Tilde{d}$ and perturbation wavenumber $q$. By definition $\log(G_r) \ge 0$.
In the drift regime, the lattice is disrupted on the time scale $t^*$.  

\subsubsection{Distinct clumping mechanisms in the exponential and algebraic growth regimes}

As explained in 
\citet{chajwa2020waves}, there are two mechanisms at play which decide the nature of clustering. The Crowley mechanism acts to form clumps at the valleys while the drift mechanism counteracts it (see figure \ref{fig:mechanisms}). 
The fast decay of growth rate with increased lattice spacing (see figure \ref{fig:decayWithD}) is because of the drift mechanism competing better against the Crowley mechanism. 
For a lattice of spheres the drift mechanism is absent, and the growth rate follows $\sigma \sim \Tilde{d}^{-2}$ for large lattice spacing $\Tilde{d}$. In contrast, for increasingly flatter oblate ellipsoids, the growth rate is best fit by $\sigma \sim \Tilde{d}^{-4.5}$ , as shown in figure \ref{fig:decayWithD}.

Our objective is to develop a statistical measure that can differentiate between clustering in the Crowley regime and the drift regime. The drift mechanism leads to aggregation in pairs whereas in the Crowley regime a larger number of discs participate in forming an aggregate, thus suggesting $P_s(t)$, the fraction of particles participating in  aggregates of size $s$ as a useful quantitative measure. Discs that are within a distance of $0.5a$ of each other are defined to be part of an aggregate. 
We conduct 500 simulations in both the Crowley and the drift regimes, each involving a lattice of 60 discs with periodic boundary conditions along the horizontal ($x$) direction.  
Each disc interacts with eight neighbors on either side.
Changing the number of neighbors about this cutoff scale does not affect cluster sizes and types, provided there are a sufficient number of discs interacting with their neighboring ones upon spatial rearrangement as the system evolves.
Each simulation is initialized with sinusoidal perturbations only in the $x$-direction, at a chosen dominant wavenumber, with amplitude $0.625a$, and additional random perturbations of amplitude $0.075 a$, where $a$ is the semi-major axis of the discs. The initial orientation angle is chosen from a uniform distribution of width $8$ degrees. The amplitude of the perturbations and the noise in initial conditions emulate the perturbations in the experiments.

\begin{figure}
\centering
\includegraphics[width=0.7\textwidth]{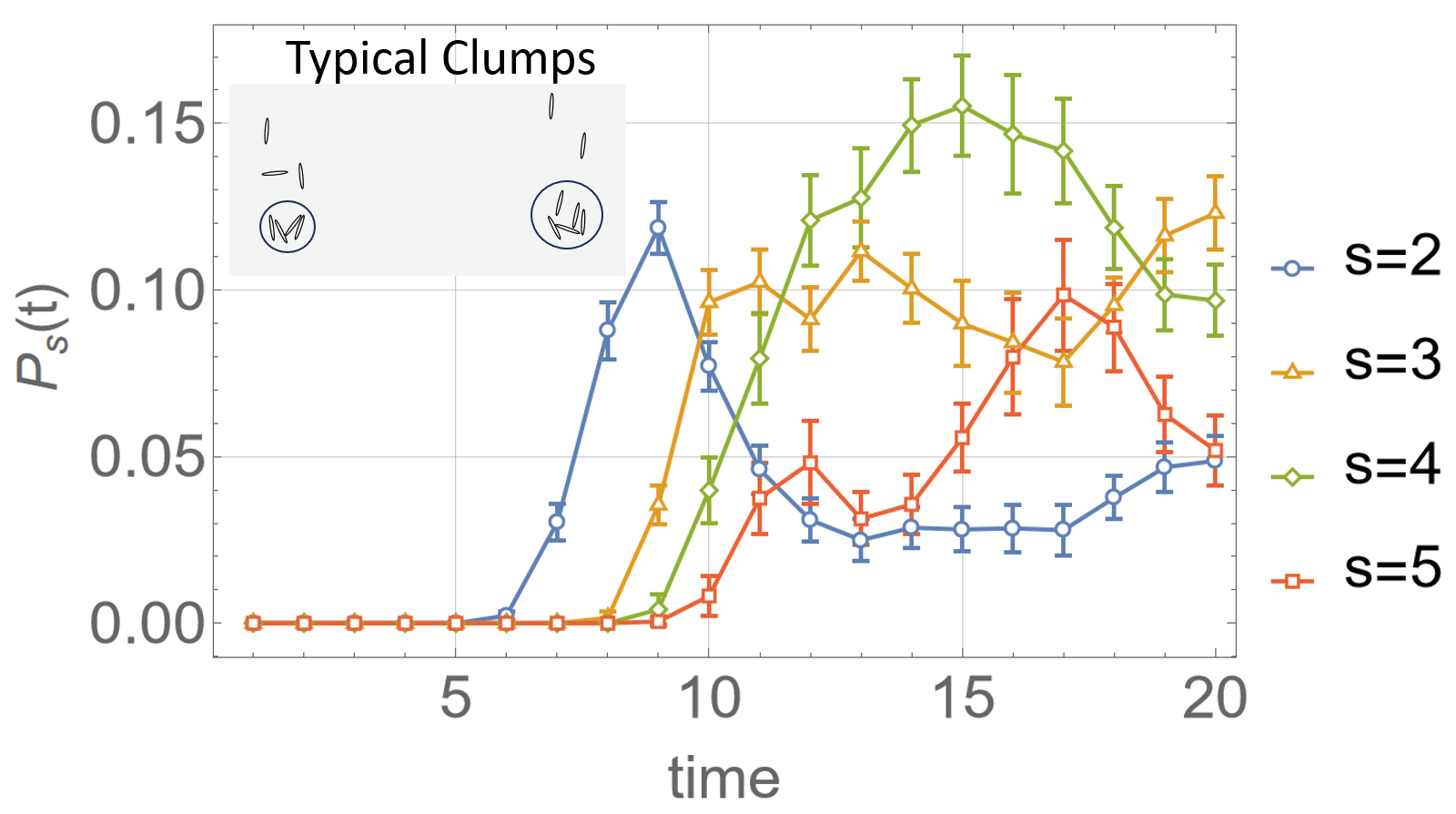}
\caption{\label{fig:histU} Probability distribution $P_s(t)$ as a function of non-dimensional time for different aggregate size $s$, in the Crowley regime $(q,\Tilde{d})=(\pi/6,1.875)$. 
$P_s(t)$ is calculated as the ensemble average of the fraction of discs forming an aggregate of size $s$ relative to the total number of discs in each ensemble. 
Discs that are within the distance of $0.5a$ are considered to be aggregated or clustered.
The error bars represent the standard deviation.
}
\end{figure}

\begin{figure}
\centering
\includegraphics[width=0.6\textwidth]{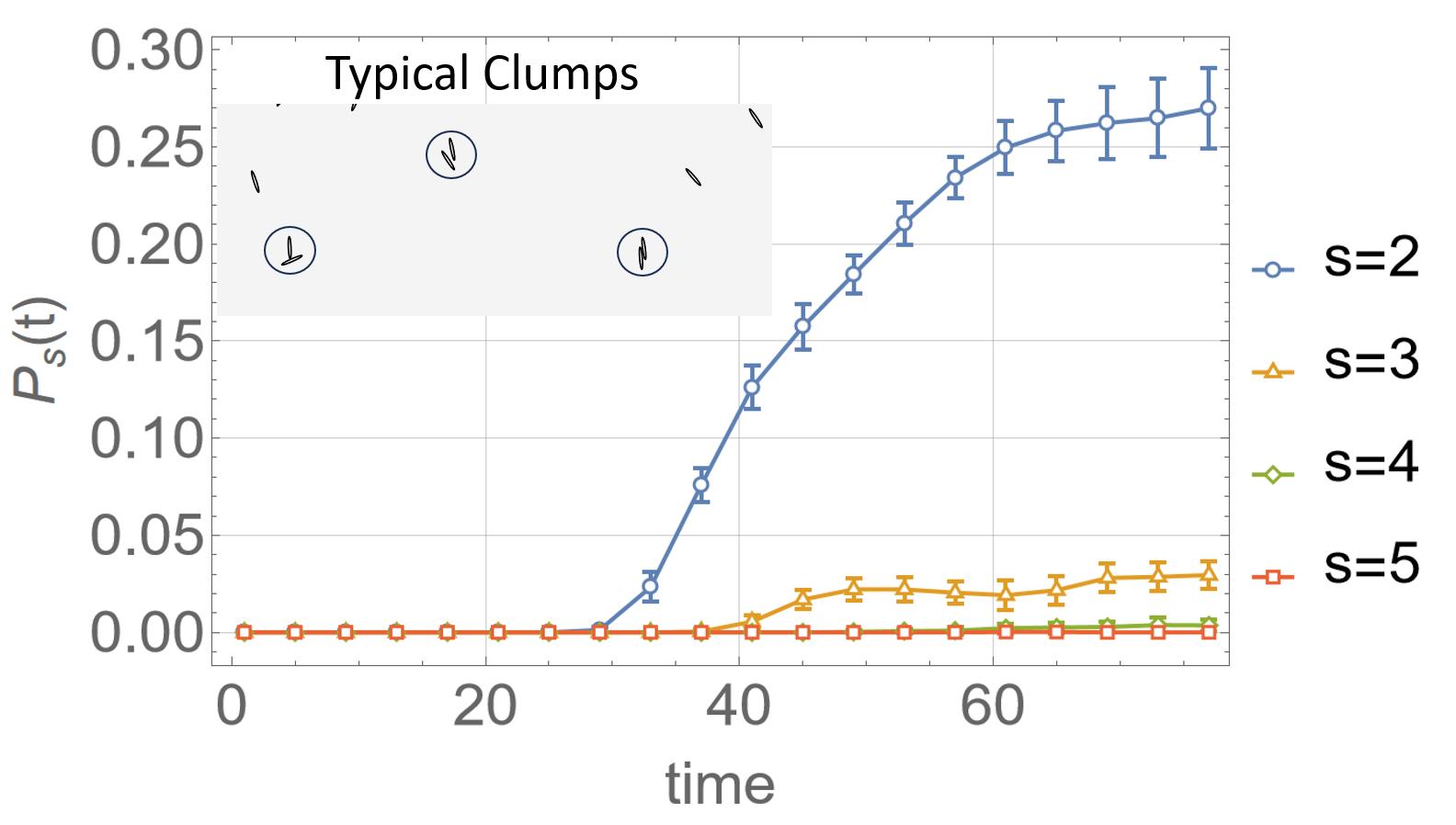}
\caption{\label{fig:histS} Probability distribution $P_s(t)$ as a function of non-dimensional time for different aggregate size $s$,  in the drift regime $(q,\Tilde{d})=(\pi/2,3.75)$. 
$P_s(t)$ is calculated as the ensemble average of the fraction of discs forming an aggregate of size $s$ relative to the total number of discs in each ensemble.
The pairs of discs separate out in a `$\perp$' configuration from the rest and keep on settling together, maintaining their orientations. 
}
\end{figure}

\begin{figure}
\centering
\includegraphics[width=0.6\textwidth]{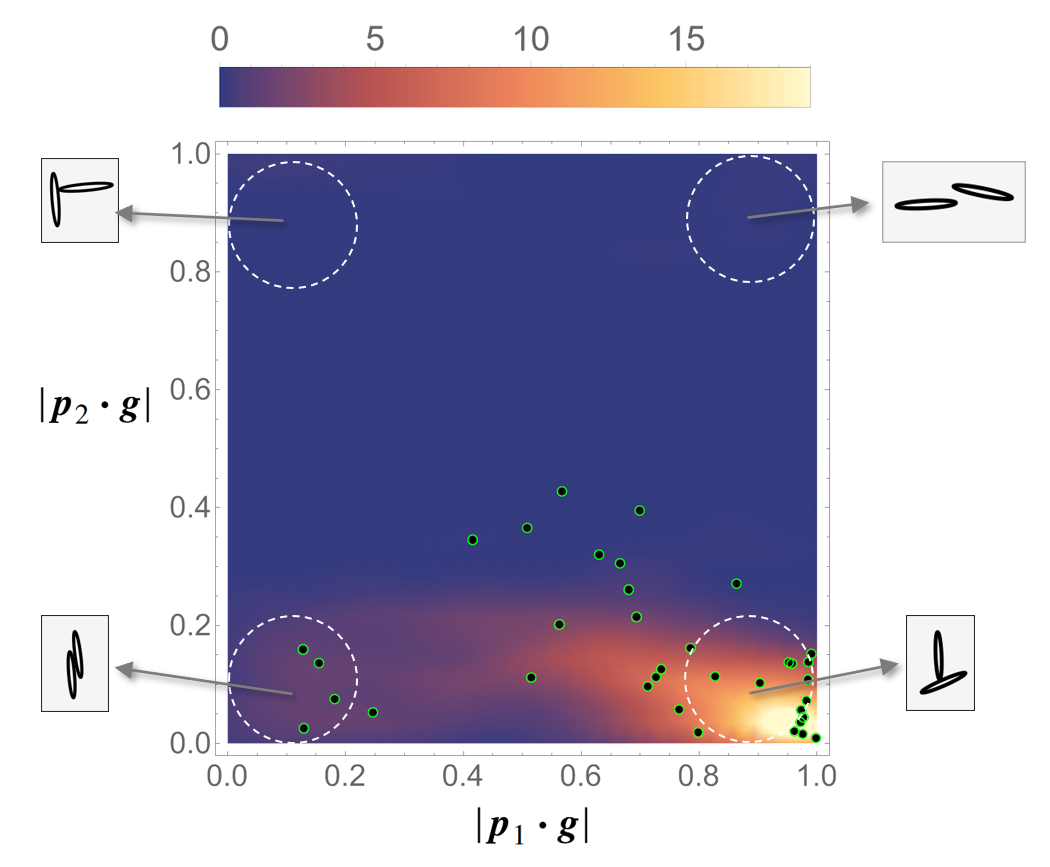}
\caption{\label{fig:pDotgS} 
Probability density plot of orientations of pairs of discs in the drift regime $(q,\Tilde{d})=(\pi/2,3.75)$. $\boldsymbol p_1$ refers to the orientation of the disc 1 which is below disc $2$ and $\boldsymbol{g}$ is the gravity unit vector. The black-green dots are experimentally observed orientations and the background colour indicates the frequency of occurrence of a given pair of orientations in the simulations. A predominance of the `$\perp$' formation is seen in both experiment and simulations. The orientation statistics are acquired when the pairs start forming in the simulations, at a non-dimensional of $t\sim 45$ . This time is later than $t^*$, the time at which the non-normal growth reaches its maximum, as shown in Figure \ref{fig:GvsExp}. The black-green dots, indicating experimental samples, are taken when pair formation is observed.
}
\end{figure}

Figure \ref{fig:histU} shows the time evolution in the Crowley regime of the fraction of discs participating in clusters of different sizes. The 500 samples are divided into 20 sub-samples, over which the mean and standard deviation are calculated.  
The most frequently occurring clump consists of four discs, while the largest clump contains five discs. Upon examination, we find no preferred relative orientations among the discs within these clumps.
A key characteristic of the drift regime is that pairing is by far the most frequent type of clustering observed. Each pair adopts a `$\perp$' shape (see figure \ref{fig:lowQClump}). This is an instance of hydrodynamic screening, as discussed in section \ref{sec:valdtn} and figure \ref{fig:compareExp}. 
With time, the paired discs separate out from the rest of the discs because they settle faster than isolated discs, usually without ever separating from each other. As a result, the fraction of pairs (or $s=2$ aggregates) only increases with time, as shown in Figure \ref{fig:histS}. 
The relative orientation of a pair of discs is quantified by  $\boldsymbol{p}_1\boldsymbol{\cdot}\boldsymbol{g}$ and $\boldsymbol{p}_2\boldsymbol{\cdot}\boldsymbol{g}$. Here, $\boldsymbol{g}$ is the gravity unit vector and $\boldsymbol{p}_1$ and $\boldsymbol{p}_2$ are respectively the orientations of the lower and upper disc of a pair. 
As previously mentioned, the discs come together in a `$\perp$' configuration, i.e. $(|\boldsymbol{p}_1\boldsymbol{\cdot} \boldsymbol{g}|,\, |\boldsymbol{p}_2\boldsymbol{\cdot} \boldsymbol{g}|)=(1,0)$ as they begin to become a pair, as evident in figure \ref{fig:pDotgS}. 
These `$\perp$' configurations, also observed in our experiments, are solely governed by two-body hydrodynamic interactions and are found to remain stable over time in numerical simulations. 
Although the experiments show the `$\perp$' signature at the time of pair formations, the pair-orientations exhibit significant variability as the discs approach the base of the container. 
Given the uncertainty in the factors influencing the persistence of `$\perp$’ structures in the experiments, we refrain from making conclusions about the late-time ($t \gtrsim 60$) dynamics and evolution of clusters based on our simulations.

\section{Formation of the `$\perp$' shape}
\label{sec:PerpShape}
The phenomenon of pairing through drift merits attention as a distinctive property of the settling of non-spherical objects, ruled out for the Stokesian settling of two spheres. As two spheroids approach each other, they invariably rearrange themselves into a $\perp$ structure, both in our experiments and in our simulations.  
This process can be explained by a simple analysis similar to that of \citet{koch1989instability} (see figure 1 in their paper). 
According to their description, the neighbors of a disc tend to align their `thin' sides parallel to the direction of the extensional component of the flow generated by that disc.
However, we point to a key difference coming from the vorticity disturbance due to the rotation of the spheroid. Consider a force monopole at the origin and a disc of radius $a$ settling in its flow field at a radial distance $r$, with position $(x,0,z)$. With contributions from both the vorticity and strain-rate of the flow produced by the monopole, it follows from equation \eqref{eq:dDotPt} that the orientation of the disc $\boldsymbol{p}=(\cos\theta,0,\sin\theta)$ evolves, to leading order in $a/r$ as
\begin{equation}
    \label{eq:thetaEqn}
    \frac{d\theta}{dt} = \frac{3}{4r^2}\bigg[\cos\phi-\zeta \sin\phi\sin 2(\phi-\theta)\bigg]; \quad \zeta\equiv \frac{3e^2}{4-2e^2}. 
\end{equation}
Here $x=r\cos\phi$ and $z=r\sin\phi$.
Equation \eqref{eq:thetaEqn} is non-dimensionalised using the same scales as described in section 4. 
Note that the first term in the right hand side of equation \eqref{eq:thetaEqn} represents the vorticity contribution and the second term represents the contribution of the strain-rate to the disc's rotation.

Now, without the vorticity contribution one gets the stable orientations seen in the schematic figure 1 of \citet{koch1989instability} and thereby an inward particle flux from all directions. In our study, both the vorticity and strain-rate contributions are crucial to the disc's rotation. To find the stable orientations while incorporating the vorticity contribution, we proceed by defining the angle
 \begin{equation} 
 \beta \equiv \frac{1}{2} \arcsin \frac{\cot\phi}{\zeta} .
 \label{eq:betaDef}
 \end{equation} 
 \begin{figure}
     \begin{center}
     \includegraphics[width=0.5\textwidth]{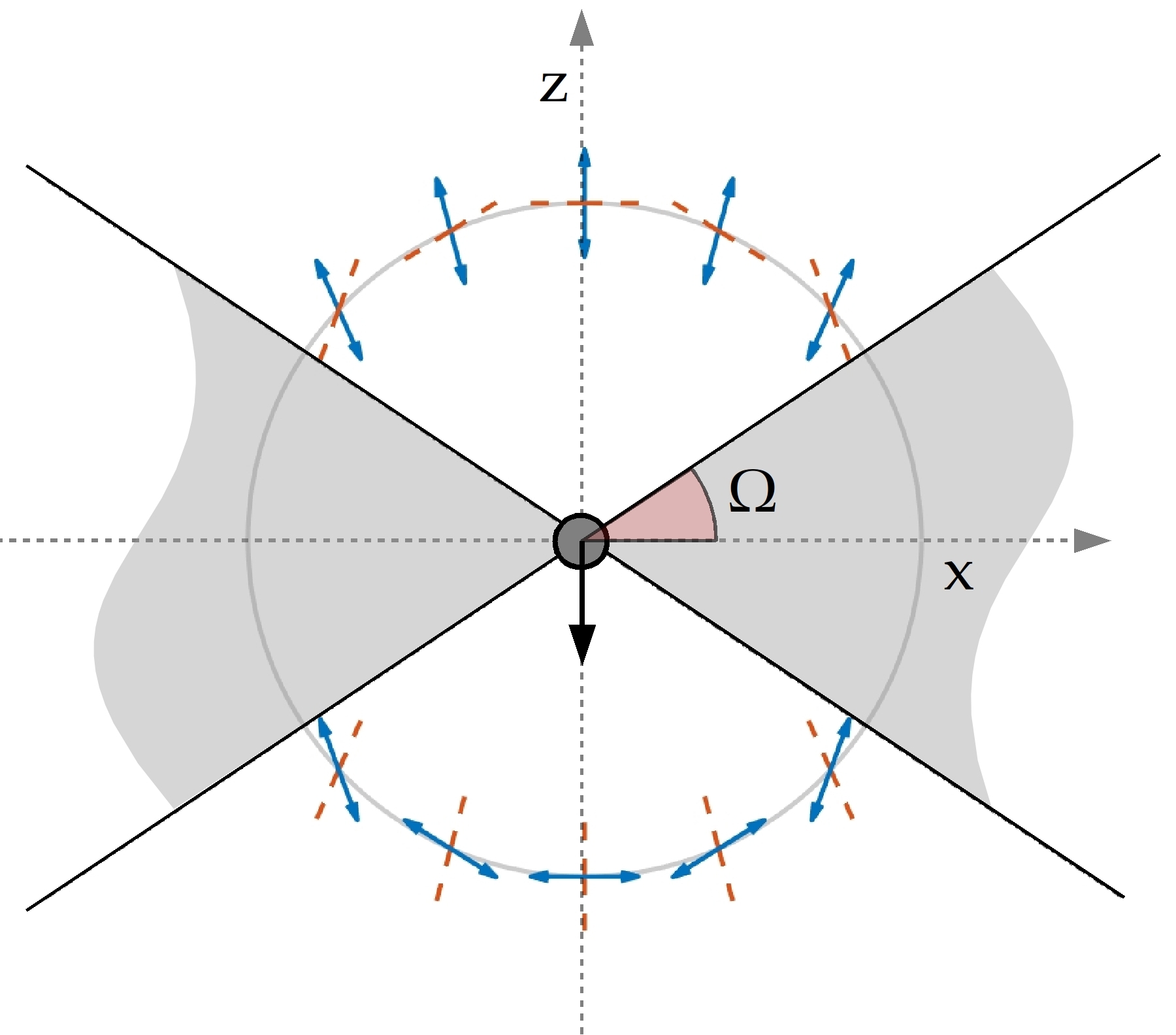}
     \caption{\label{FigKSmod} Stable orientations with both vorticity and strain-rate contributing to a disc's rotation: with a force monopole at the origin. Blue double-ended arrows and red-dashed lines are stable and unstable alignments of the thin side of the disc respectively. The  angle between stable and unstable orientations is $\pi/2 - 2\beta$. The shaded grey region represents the range of angles $\phi$ where there are no stable orientations, with $\Omega = \arctan(1/\zeta)$. }
     \end{center}
     \end{figure}
For any 
$(r,\phi)$, we set $d\theta/dt=0$ in \eqref{eq:thetaEqn} to get the fixed point
\begin{equation}
    \label{eq:thetaFP}
    \theta^* = \phi - \frac{n\pi}{2}-(-1)^n\beta, \quad n\in \mathbb{Z}
\end{equation} 
for the orientation angle $\theta$.
We can immediately see from \eqref{eq:thetaFP} that solutions do not exist for all values of $\phi$, which is a qualitatively distinct feature from the picture of \citet{koch1989instability}.  For $\Omega \equiv \arctan(1/\zeta)$, the solution for $d\theta/dt =0$ is defined only for the values of $\phi$ in the intervals $(\Omega , \pi - \Omega) $ and $(\pi + \Omega, 2\pi - \Omega) $. Even where the solution exists we show below that the stable orientations are different from what one gets with just the strain-rate contribution. 
 
We can study the stability by considering just the solutions $\theta^*_1 = \phi - \beta$ and $\theta^*_2 = \phi + \beta - \pi/2$, given the apolarity of discs. Perturbing the fixed points by small angles $\delta \theta_1$ and $\delta \theta_2$ respectively, substituting in \eqref{eq:thetaEqn} and Taylor expanding about the respective fixed points gives
\begin{equation}
\dot{\delta \theta}_k = (-1)^{k-1} \cos 2 \beta \, \frac{ 3 \zeta \sin \phi}{2 \, {r^{2}}}\, \, \delta \theta_k , \quad k\in \{1,2\}.
\end{equation}
Since $\cos 2\beta>0$, the fixed point $\theta^*_1$ is unstable in the upper half plane and stable in the lower half plane and vice versa for the other fixed point $\theta^*_2$. 

To summarize, in the upper half-plane (when the disk is above the force monopole) the disc is stable with its orientation along the angle $\theta_2^* = \phi - \beta$. In the lower half-plane the stable angle of the orientation is $\theta_1^* = \phi + \beta - \pi/2$. These stable orientations are shown in figure \ref{FigKSmod} with blue arrows indicating the stable alignment of the thin side of the disc \footnote{Note that the thin side of the disc is perpendicular to its orientation vector}. In both half-planes, solutions exist only with the azimuthal angle $\phi$ lying in the intervals $(\Omega , \pi - \Omega) $ and $(\pi + \Omega, 2\pi - \Omega) $, leaving out regimes shown as grey shaded region in figure \ref{FigKSmod}, where there are no stable orientations.

With this picture in mind we can explain the formation of `$\perp$' shapes as follows. For a pair of vertically separated settling discs, the force monopole flow created by the lower disc aligns the upper disc vertically, and that generated by the upper disc aligns the lower disc horizontally.
Since a vertically aligned disc sediments more rapidly than a horizontal one, the upper disc catches up with the lower one,  forming the `$\perp$' shape, frequently observed in disc pairs. In the drift regime the lattice spacing  is large enough, i.e., the system is dilute enough, for the above approximate analysis based on pair interactions and point-force flows to suffice.
The absence of this `$\perp$' shape in the Crowley regime indicates that clusters in this regime are influenced by multi-particle interactions.

\section{Conclusions and suggestions for future work}
The slow collective sedimentation of discs is governed by two processes -- the Crowley mechanism and orientation-dependent drift -- whose competition was observed and the resulting stability problem studied in \citet{chajwa2020waves}.
We have explored how these mechanisms lead to disruption of a settling lattice of discs, creating two kinds of structures: clumps of several particles and pairs which remain together forever respectively.  
We have shown that a satisfactory understanding of the near-contact dynamics of a pair of discs crucially requires the inclusion of the first-reflection contribution to their hydrodynamic interaction.
We then analyzed the stability of a one-dimensional lattice of discs using the first reflection and contrasted it with the point force approximation used in the earlier work of \citet{chajwa2020waves}. Whereas the point force approximation predicts zero growth rate $\sigma$ beyond a critical lattice spacing $\Tilde{d}_c$, the first reflection shows that the configuration is unstable at all spacings, but is extremely weak at large lattice spacing $\Tilde{d}$, with: $\sigma\sim \Tilde{d}^{-4.5}$, a much faster decay than the case of spheres. Nevertheless, there are two regimes based on the whether or not the Crowley mechanism overpowers the drift mechanism, which can be quantified by looking at the nature of clustering. 
In terms of perturbation growth, the Crowley regime leads to exponential growth while the drift regime shows transient algebraic growth, both of  which eventually lead to the disruption of the lattice.   
The Crowley mechanism leads to clumping at the valleys of the disrupted lattice while in the drift regime discs cluster in pairs, away from the valleys, forming `$\perp$' structures. 
These `$\perp$' configuration 
exemplify a type of hydrodynamic screening or shadowing, in which
the upper disc descends vertically onto the lower disc, which has its flat side facing downward.
These `$\perp$' structures are observed in both isolated disc pair experiments and one-dimensional disc lattices. In the isolated disc pair experiments, they demonstrate high stability and retain their configuration for an extended period.
We have explained these structures through an analysis that builds on \citet{koch1989instability}'s work by incorporating vorticity contributions to the rotation of the discs.

So far, we have seen how the clustering of a one-dimensional array of sedimenting spheroids is fundamentally different from spheres. Several directions for future investigation suggest themselves, and we mention a few. Studying the sedimentation of a specific arrangement provides a useful test bed for understanding how hydrodynamic effects operate at both single and multi-particle levels. But practical scenarios, like the settling of centric (\textit{Coscinodiscus}) and pennate (\textit{Pseudo-nitzschia}) diatoms in an algal bloom, often involve homogeneous suspensions of oriented particles, with clustering being crucial for `marine snow' formation \citep{Burd2009}. We may ask how clustering depends on the initial spatial distribution of non-spherical particles, especially in two and three dimensions. Secondly, our work has been restricted to the steady Stokes (zero Stokes number) limit. When the background fluid is turbulent, inertial particles, i.e., particles of non-zero Stokes number, are well known to form clusters even in the dilute limit where inter-particle interactions are absent, see e.g. \citet{bec2003fractal,monchaux2010preferential,reade2000effect}. Our study indicates that a future theory which includes interparticle interactions in unsteady Stokes flow (finite Stokes number) is in order. During sedimentation a single spheroid can only execute persistent drift at a constant speed. But this behaviour is seen only in a small fraction of anisotropic particles \citep{PhysRevLett.134.014002}, which can exhibit settling, drifting or quasiperiodic motion \citep{miara2024dynamics, witten2020review, GONZALEZ_GRAF_MADDOCKS_2004, makino2005sedimentation, tozzi2011settling, krapf2009chiral}. 
Understanding how departures from spheroidal shape combines with hydrodynamic interactions among particles will yield interesting results relevant to ice formation in clouds and the clumping of marine snow.

\backsection[Funding]{We acknowledge support of the Department of Atomic Energy, Government of India, under project no. RTI4001. NM was supported through NSF DMR 2319881. RC acknowledges support from the International Human Frontier Science Program Organization. }

\backsection[Declaration of interests]{The authors report no conflict of interest.}



\appendix
\section{Finding points of closest approach  between two discs}

\begin{figure}
\centering
\includegraphics[width=0.8\textwidth]{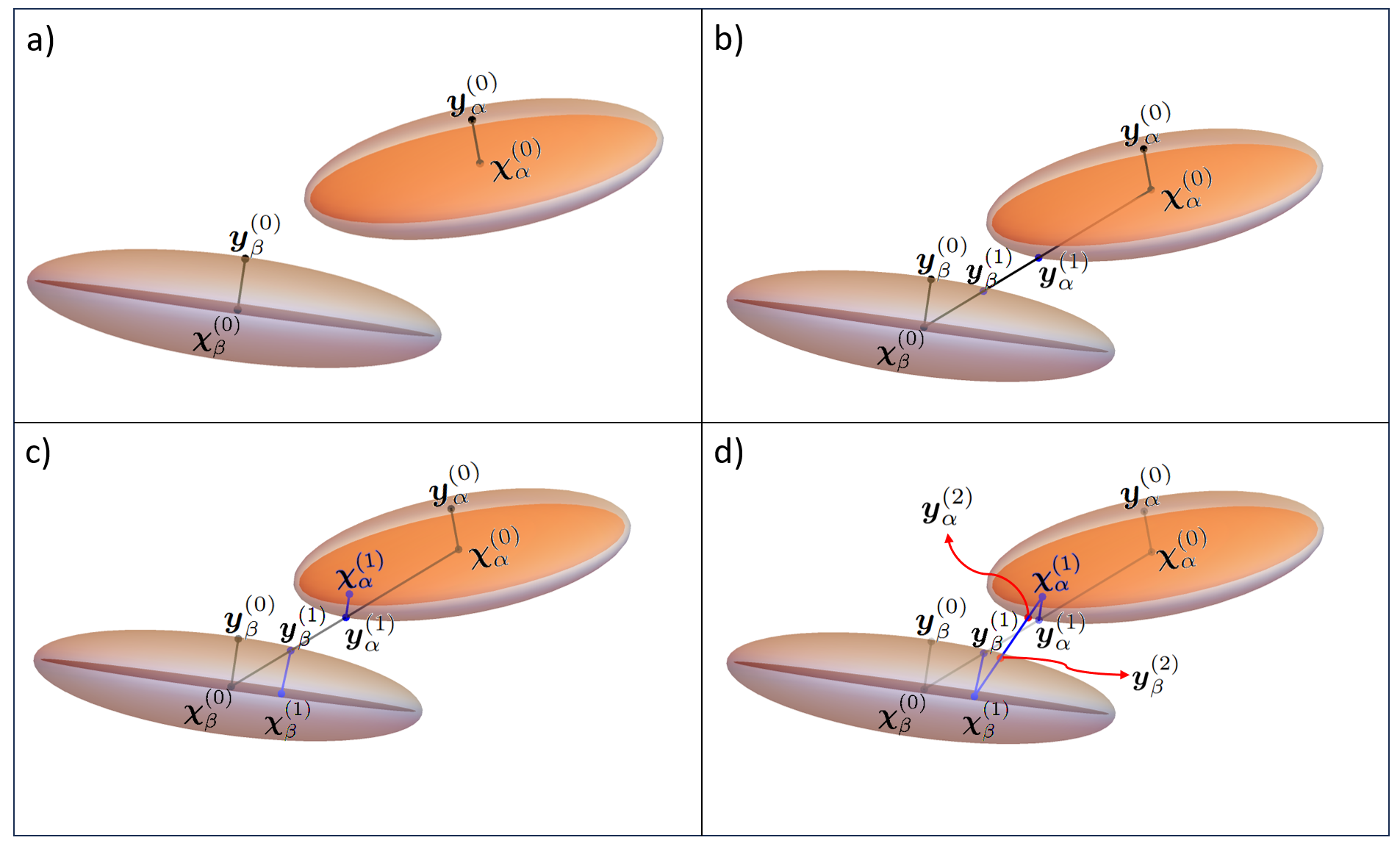}
\caption{\label{fig:collsnDetectn}Choosing any points $\boldsymbol y_\alpha^{(0)}$ and $\boldsymbol y_\beta^{(0)}$ on the discs, one can reach the closest points iteratively by following the algorithm described in the text. The labels a) to d) shows the candidate points obtained for the closest points after two iterations. The dark plane in the figures defines the confocal disc at the midplane of the oblate spheroid.}
\end{figure}

This section describes the algorithm to find the minimum distance between two spheroids. We extend the algorithm of \citet{claeys1993suspensions} 
for prolate spheroids to the case of a pair of oblate spheroids (discs). 
The main idea is to exploit the fact that the outward normal vectors of two spheroids at their points of closest distance are anti-parallel. Any point $\boldsymbol x$ on the surface of the disc $\alpha$ can be defined by 
\begin{equation}
    \label{eq:ellipse1}
     (\boldsymbol{x} - \boldsymbol{x}_{\alpha} )^{T}\boldsymbol{\cdot} \boldsymbol\Sigma_\alpha \boldsymbol{\cdot} (\boldsymbol{x} - \boldsymbol{x}_{\alpha} ) = 1,
\end{equation}
where
$\boldsymbol x_{\alpha}$ is the centre of the disc $\alpha$, and
\begin{equation}
    \label{eq:wgtMat}
    \boldsymbol\Sigma_\alpha = \left[ \frac{\bp_\alpha \bp_\alpha}{b^2} + \frac{(\boldsymbol{\delta}-\bp_\alpha\bp_\alpha)}{a^2} \right] 
\end{equation}
is known as the weight matrix with $b=a\sqrt{1-e^2}$.
Taking the gradient of \eqref{eq:ellipse1} gives us the normal vector to the disc at $\boldsymbol x$:
\begin{equation}
    \label{eq:normalEllipse1}
    \boldsymbol n_\alpha(\boldsymbol x) =  \frac{\boldsymbol\Sigma_\alpha \boldsymbol{\cdot} (\boldsymbol{x} - \boldsymbol{x}_{\alpha} )}{|\boldsymbol\Sigma_\alpha \boldsymbol{\cdot} (\boldsymbol{x} - \boldsymbol{x}_{\alpha} )|}.
\end{equation}
We start with two points $\boldsymbol y_\alpha^{(0)}$ and $\boldsymbol y_\beta^{(0)}$ on discs $\alpha$ and $\beta$ respectively, and construct a sequence of points $\boldsymbol y_\alpha^{(k)}$ and $\boldsymbol y_\beta^{(k)}$ following the iterative procedure outlined below to converge rapidly to the points of closest distance between discs $\alpha$ and $\beta$:
\begin{itemize}
    \item Construct a normal vector $\boldsymbol n_\alpha$ at $\boldsymbol y_\alpha^{(k)}$. Find the point of intersection of the line along $\boldsymbol n_\alpha$ passing through $\boldsymbol y_\alpha^{(k)}$ and the mid-plane of the oblate spheroid $\alpha$ (see figure \ref{fig:collsnDetectn}). Call this point $\boldsymbol \chi_\alpha ^{(k)}$. Repeat for the disc $\beta$.
    \item Construct a line joining $\boldsymbol\chi_\alpha^{(k)}$ and $\boldsymbol\chi_\beta^{(k)}$. The points of intersection of this line and the two discs give the next iterates $\boldsymbol y_\alpha^{(k+1)}$ and $\boldsymbol y_\beta^{(k+1)}$.
\end{itemize}

Thus, one can find the minimum separation vector $\boldsymbol\epsilon_{\alpha\beta}$ as:
\begin{equation}
    \label{eq:minD}
    \boldsymbol\epsilon_{\alpha\beta} = \boldsymbol y_\alpha - \boldsymbol y_\beta,\quad \boldsymbol y_\alpha = \lim_{k\to \infty} \boldsymbol y_\alpha^{(k)},\quad \boldsymbol y_\beta = \lim_{k\to \infty} \boldsymbol y_\beta^{(k)}.
\end{equation}
The numerical value of the closest distance can be obtained by repeating the two steps described above until the current and the previous iterations differ by some small tolerance value. The algorithm described above converges to the points of closest distance when the discs are not overlapping, with a minimum separation vector $\boldsymbol{\epsilon}$.

\bibliographystyle{jfm}
\bibliography{jfm}


\end{document}